\documentclass{elsart}

\usepackage{graphicx}
\usepackage{epsfig}

\usepackage{amssymb}
\usepackage{color}

\begin{document}

\begin{frontmatter}



\title{A Comprehensive Study of Low-Energy Response for Xenon-Based Dark Matter Experiments}
\author[usd]{L.Wang},
\author[usd,yzu]{D.-M. Mei\corauthref{cor}}
\corauth[cor]{Corresponding author.}
\ead{Dongming.Mei@usd.edu}

\address[usd]{Department of Physics, The University of South Dakota, Vermillion, South Dakota 57069}
\address[yzu]{School of Physics and Optoelectronic, Yangtze University, Jingzhou 434023, China}

\begin{abstract}
\label{abstract1}
We report a comprehensive study of the energy response to low-energy recoils in dual-phase xenon-based dark matter 
experiments. A recombination model is developed to explain 
the recombination probability as a function of recoil energy at zero field and non-zero field.
The role of e-ion recombination is discussed for both parent recombination
and volume recombination. We find that the volume recombination under non-zero field is constrained by a plasma effect,
which is caused by a high density of charge carriers along the ionization track forming a plasma-like cloud of charge 
that shields the interior from the influence of the external electric field. Subsequently, the plasma time that determines
the volume recombination probability at non-zero field  
is demonstrated to be different between electronic recoils and nuclear
recoils due to the difference of ionization density between two processes.
We show a weak field-dependence of the plasma time for nuclear recoils and a stronger field-dependence of the
plasma time for electronic recoils. 
As a result, the time-dependent recombination is 
implemented in the determination of charge and light yield with a generic model. 
Our model agrees well with the available experimental data from xenon-based dark matter experiments. 
\end{abstract}
\end{frontmatter}

\maketitle

\section{Introduction}
\label{Intro}
Observations commencing in the 1930s~\cite{fzw} have led to the contemporary and rather shocking view that 96\% of the matter and energy in the universe 
neither emits nor absorbs light or other electromagnetic radiation~\cite{chi, rjg}.  The most popular candidate for the dark matter is the WIMP (Weakly 
Interacting Massive Particle), a particle with a mass thought to be comparable to the mass of heavy nuclei, but with feeble and extremely short-range
(in the weak-force scale) 
interaction with atomic nuclei. Theories invoking Supersymmetry (SUSY), being probed currently at the Large Hadron Collider (LHC), 
naturally provide a particle that could be the WIMP~\cite{jlf}. 
Most of the mass of the Milky Way galaxy would be WIMPs, and they would interact with matter so rarely that WIMPs would easily pass through the Earth. 
 Very rarely, WIMPs would collide with atomic nuclei, causing the atom to suddenly recoil with a velocity thousands of times that of sound~\cite{mwg}.
 Many studies over the past decade have emphasized the high priority of WIMP direct detection~\cite{mtu, ppp}. 
Over thirty years exploration with many
targets, liquid
xenon has become a leading technology in the field. The LUX (Large Underground Xenon), XENON100, 
and PandaX~\cite{luxd, LUX, xenon100, pandax} experiments have 
demonstrated the scalability of xenon detector to multi-ton scale. It is the goal of 
the successor experiments, LZ (LUX-Zeplin) and XENON1T~\cite{lz, xenon1t} to decisively identify xenon atoms that have suddenly recoiled 
in response to a collision with a WIMP.  
Successful identification (known as direct detection) and subsequent studies would transform empirical understanding of cosmology, 
astrophysics, and particle physics.

\begin{figure}[htb!!]
\begin{center}
\includegraphics[angle=0,width=12.cm] {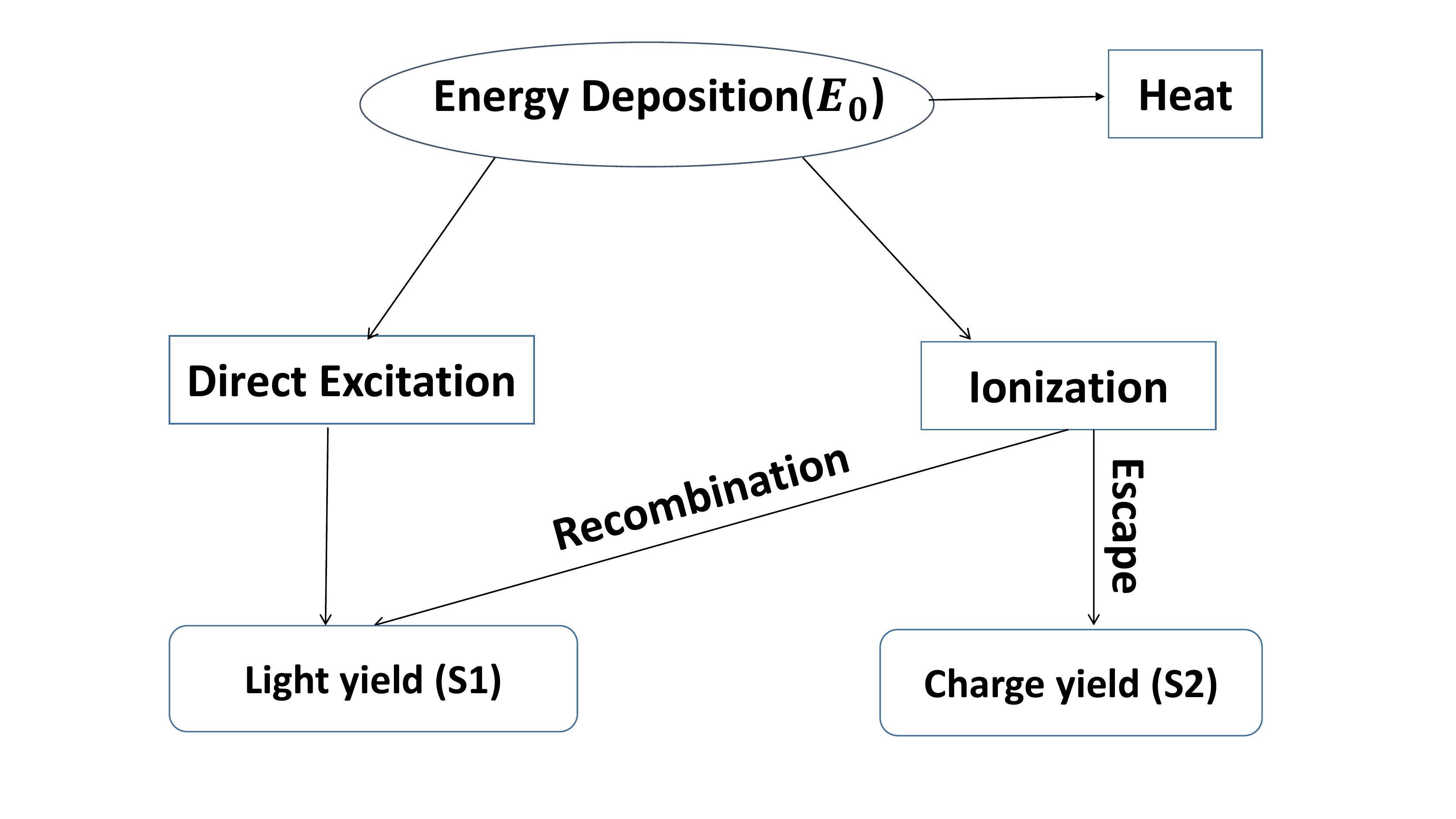}
\caption{\small{A schematic energy deposition in dual-phase xenon detectors.}}
\label{fig:sche}
\end{center}
\end{figure}

The detection of WIMP-induced nuclear recoils (NRs) in a dual-phase xenon detector can be conducted through observation of two anti-correlated and complementary signals: the
scintillation photons in liquid and the charge carriers in gas~\cite{case, ekp, eap}. 
It has been shown that
the ratio of the charge to light signal can be used to discriminate NRs against electronic recoils (ERs)~\cite{luxd, LUX, xenon100, pandax, xe10}.
The energy reconstruction in a dual-phase xenon detector is conducted based on the physics processes involved. 
An incoming particle with a given kinetic energy, interacts with xenon atoms
by depositing energy ($E_{0}$) in liquid xenon through many processes, as sketched in Figure~\ref{fig:sche}. 
The total energy deposited is expended in the production of excitons, e-ion pairs and atomic motion (heat). 
The Platzman equation~\cite{Platz} is used to describe the conservation of energy:  
\begin{equation}
E_{0} = N_{ex}\times E_{ex}+N_{i}\times E_{i}+N_{i}\times\varepsilon,
\label{eq:plasma}
\end{equation}
where $N_{i}$ is the number of e-ion pairs produced at the average expenditure of energy $E_{i}$,
$N_{ex}$ is the number of excitons produced at the average expenditure of energy $E_{ex}$, and $\varepsilon$ is the average kinetic 
energy of sub-excitation of electrons. 

The energy deposition processes create direct excitation and ionization. In a dual-phase detector, 
under an electric field, some of the electrons generated through ionization in the liquid are drifted to the gas. Therefore,
two signals are measured. The first is the primary scintillation light due to 
direct excitation and recombination of e-ion pairs in liquid, denoted as $S1$. The second is the proportional scintillation 
light in gas, denoted as $S2$, which is proportional to the number of electrons escaping the recombination process.   
One can express $<S1> = g_{1}n_{\gamma}$ and $<S2> = g_{2}n_{e}$, where  n$_{\gamma}$ is the number 
of detected photons in liquid, n$_{e}$ is the number 
of electrons detected in the gas phase~\cite{Akerib}, and $g_{1}$ and $g_{2}$ are the detector specific gain factors in liquid and gas xenon, respectively.

Since n$_{e}$ and n$_{\gamma}$ are anti-correlated with the recombination probability between electrons and ions in liquid, we can write as a generic model: 
\begin{equation}
n_{e} = N_{i}(1 - r) =  \frac{L\cdot E_{0}}{W_{i}}(1-r),
\label{eq:number1}
\end{equation}
and 
\begin{equation}
n_{\gamma} = N_{ex} + rN_{i} = \frac{L\cdot E_{0}}{W_{i}}(r+\frac{N_{ex}}{N_{i}}),
\label{eq:number2}
\end{equation}
where $r$ represents the recombination probability between electrons and ions, $L$ is the quenching factor~\cite{lind} that has
different forms depending on the recoil type, non-zero field and zero field, the $W_{i}$-value is the average energy expended per e-ion pair,
and $\frac{N_{ex}}{N_{i}}$ is the exciton-to-ion ratio. The
charge yield is defined as $Q_{y}$ = $\frac{n_{e}}{E_{0}}$ while the light yield is defined as $L_{y}$ = $\frac{n_{\gamma}}{E_{0}}$. 
 
To reconstruct
the recoil energy, one must know the $W_{i}$-value, the quenching factors, and the recombination probability
for a given electric field. The recent measurements from LUX~\cite{Akerib}, with a D-D neutron generator, 
together with the
available data from other experiments at different fields~\cite{xe100ap, sor1, sor2, xe10e, manz},
have shown an anti-correlation 
between the charge and light yield for NRs with the following
characteristics: 
(1) the charge yield increases as recoil energy decreases and (2) the charge yield is weakly dependent on electric field and 
varies in a relatively narrow 
band over a large range of electric field from 100 V/cm to 4000 V/cm. 
The physics mechanisms that are responsible for the observations in the charge yield are not clear in terms of the variation 
of the $W_{i}$-value and recombination. Although the $W_{i}$-value is typically treated as a constant, 
in this work we treat it as energy dependent. 
However, since its variation is small~\cite{Taka}, recombination is, thus, responsible
for the observed behaviors. The question becomes what physics mechanisms in the recombination processes cause the observations
and why the recombination probability is weakly dependent on the field strength for NRs. 

Similarly, the charge yield of ERs with the tritium data from LUX~\cite{luxtr} and the data 
from~\cite{dyu} have shown a distribution that widely spreads over a
range from $\sim$ 20 electrons/keV to $\sim$ 50 electrons/keV depending on recoil energy. In addition, the charge yield has a 
stronger dependence on
the applied electric field in comparison with the charge yield of NRs. The question is what causes such a behavior.

In the case of zero field, the observed light yield 
is usually interpreted as a relative scintillation efficiency with respect to a known gamma-ray energy, such as 122 keV
from $^{57}$Co calibration source~\cite{luxd, xenon100, xe10}. The available data from
various experiments~\cite{arne,akim,apri,plan,chep,apri9} shows a rather broad distribution, which deserves a good physics model to justify the observed data points.
The observed relative scintillation efficiency can be explained by three physics functions: the variation of the $W_{i}$-value, 
quenching factors, and 
recombination. We can once again assume that the variation of the $W_{i}$-value is small, quenching factors and recombination 
are responsible for the observed distribution. The question is which one is dominant.

Prior to this work, many authors have modeled ionization, scintillation, and recoil tracks
in xenon detectors~\cite{sor2, taj, ahp, mei, pso1, psc, fbf, maz1, maz2, eap1, jmo, mfo, wmu1, wmu2}.
A global analysis of light and charge yield in liquid xenon offers a data-driven model, which has
been popular in the field~\cite{NEST}. In this paper, a comprehensive study of physics mechanisms behind ionization, scintillation,
and recoil tracks is discussed. We develop a
model in Section 2 to demonstrate the physics mechanisms that govern e-ion
recombination for non-zero field and zero field. Subsequently, we implement
the model and compare with the available experimental data in Section 3. 
Section 4 summarizes the significant findings. Section 5 concludes the prediction
power and physics implications. 

\section{Recombination of e-ion pairs}
\label{LEc}
  
\subsection{Recombination model}

The light yield is the total effect of direct excitation of xenon atoms  and recombination of e-ion pairs created by ionization~\cite{Kubota, Dolg}. 
The recombination probability 
describes the fraction of electrons that recombine in liquids. Currently,
two recombination models are used to calculate the recombination probability:  
Birks-Doke Law is used for long range ionization tracks (for recoil energy $\ge$ 10 keV)~\cite{NiKai,Toke}; 
the Thomas-Imel model is adopted for short range ionization tracks (for recoil energy $\le$ 10 keV)~\cite{Dahl, PS, Thomas}.

\begin{figure}[htb!!]
\begin{center}
\includegraphics[angle=0,width=12.cm] {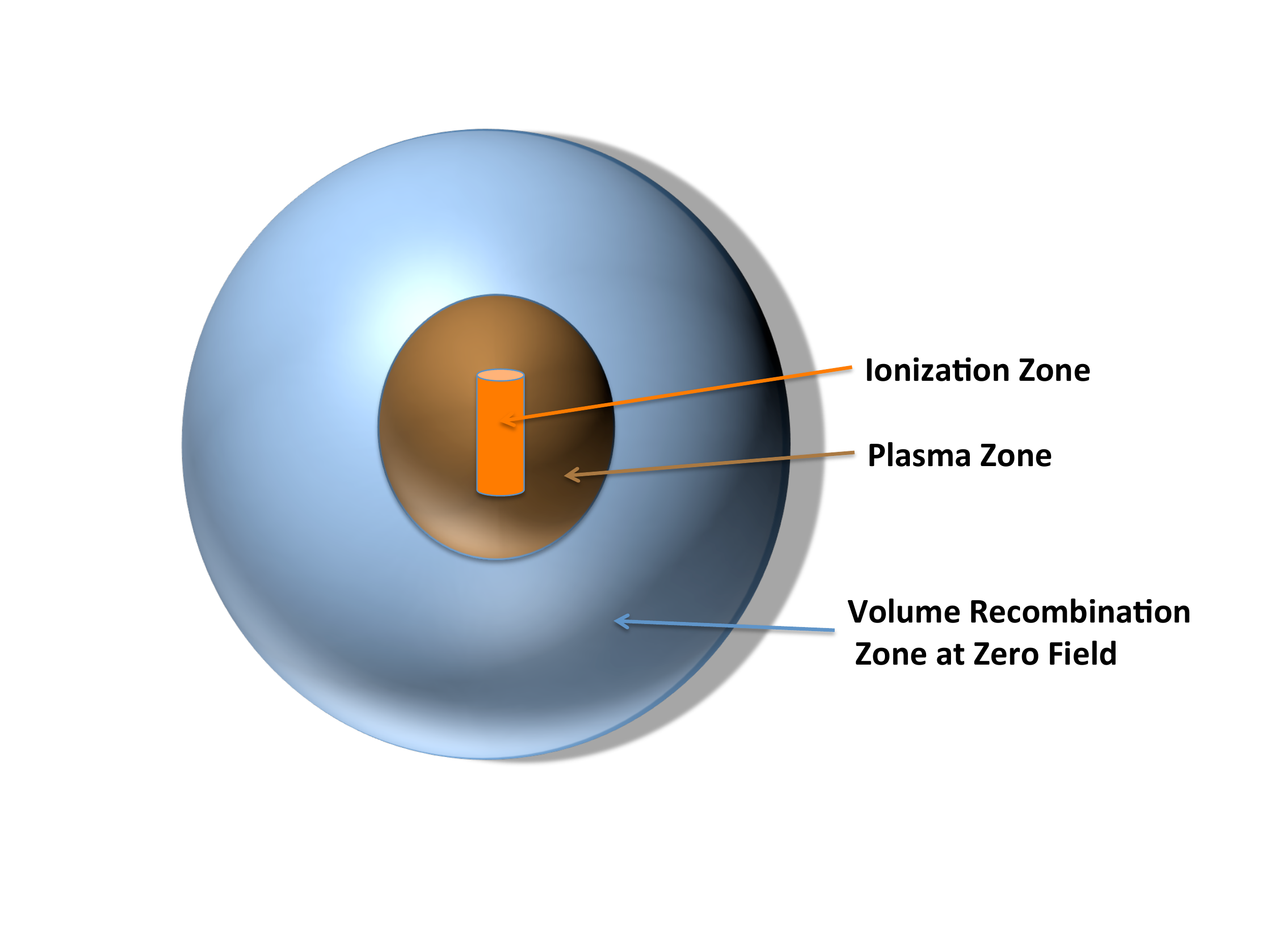}
\caption{\small{A schematic for the recombination occurring in the ionization zone, the plasma zone, and the volume recombination zone.  The ionization 
zone is assumed to be a cylinder while both the plasma zone (non-zero field) and the volume recombination zone (zero field) 
are considered as spherical shape.}}
\label{fig:zone}
\end{center}
\end{figure}

In this work, we offer an alternative physics model to describe recombination occurring under certain circumstances 
with respect to different physics processes. 
For a given energy, the charged particles ionizing atoms along their tracks create e-ion pairs.
The ions form rare-gas molecular ions, Xe$_{2}^{+}$, within picoseconds~\cite{kss, sdd, pgl}. Those molecular ions
are localized with an average distance of about 40 nm in liquid xenon for 1 MeV electrons~\cite{Kubota}. The primary and secondary
electrons are all thermalized along the track. The thermalized
electrons recombine with localized Xe$_{2}^{+}$ through formation of the excited molecular states. 
This suggests that the recombination of e-ion pairs is confined within   
a specific volume. Within this volume, recombination can be described as two stages with respect to
the recombination time scale. The first stage occurs when electrons are thermalized along the ionization track. 
These thermalized electrons recombine with their own parent ions merely 
in the ionization zone. This process is called ``the parent recombination''~\cite{Toke,Thomas}. In the second stage, 
the ionization zone is expanded outwards to be a larger zone due to the ambipolar diffusion~\cite{kiz} of electrons and ions. 
Within this larger zone, thermal electrons recombine with 
ions other than their own parent ions. This process is named as ``the volume recombination''~\cite{Toke,Thomas}.
We illustrate a two-zone recombination in
Figure~\ref{fig:zone}.

In the ionization zone, thermal electrons are mainly attracted by their own parent ions 
along the initial ionization track~\cite{Ferdor,Seibt},
and recombine with their own parent molecular
ions. The rate of the parent recombination is governed by the parent recombination time,
which is independent of electric field. 
Many previous works employed the Onsager radius in which the parent recombination takes place~\cite{MSZY, Dahl, Thomas, Onsager, SAM, Nygr}. 
However, the time scale of the parent recombination within
the region has not been used in the calculation of the recombination probability.     
Beyond the ionization zone, 
the recombination time of e-ion pairs can be very different for non-zero field and zero field. This is  
because the velocity and the directions of thermal electrons can be 
greatly affected by the presence of an external electric field.

The plasma zone is the expansion of the ionization zone in which 
a high density of charged carriers along the ionization track forms a plasma-like cloud of charge 
that shields the interior 
from the influence of the external electric field~\cite{Dolg}. Only those charge carriers arriving at the outer edge of 
the cloud are subject
 to the influence of the external electric field, and begin to 
migrate immediately. This plasma-like cloud expands radially due to the 
ambipolar diffusion of charge carriers 
and is gradually eroded away until the charges at
the interior are finally subject to the external field and also begin to drift. The time needed for the total disintegration 
of this plasma region is called  
``the plasma time'' and is responsible for the volume recombination at non-zero field. Once the plasma region disappears, 
the charge carriers are drifted away by 
the external electric field, and the recombination ceases between electrons
and ions~\cite{Dolg}. 

In the absence of an electric field, the ionization zone is expanded due to the ambipolar diffusion process as well. The electrons at the edge of the ionization zone exchange momentum and energy with xenon atoms in the medium, which is controlled by the deformation potential scattering due to the longitudinal acoustic waves~\cite{Basak}.  This is because the accompanying acoustic waves generation is required by the momentum conservation in the process of inelastic collisions between electrons and xenon atoms. As the electrons lose energy, the magnitude of acoustic waves generated by the inelastic collision process becomes weaker. As a result, the impact of the deformation potential due to the acoustic waves on the electrons becomes smaller, until the velocity of the electrons is equal to the longitudinal sound velocity, $0.65 \times 10^{5}$ cm/s~\cite{sowa} in liquid xenon, and the deformation potential of the acoustic waves disappears, hence the electrons escape recombination. The time that electrons spend to reach the longitudinal sound velocity is called ``the volume recombination time'' at zero field.

In the above two scenarios regarding non-zero field and zero field, the volume recombination probability is governed by the volume recombination time.
 At non-zero field, the volume recombination time is equivalent to the plasma time. 
Combining the parent recombination and the volume recombination together, it is clear that   
the recombination of e-ion pairs relies on the parent recombination time and the volume recombination time. 
At time $t$ = 0, the recombination probability is 0 and the recombination begins. When $t$ approaches to infinite, 
the recombination probability approaches to 1 at zero field with a sufficient large detector.

Therefore, the recombination probability, $r$, representing the probability of recombining in between t = 0 and t = t, 
can be postulated as 
\begin{equation}
r = 1-e^{-\lambda t}.
\label{eq:rec}
\end{equation}

The term $\lambda$ is recombination constant for both the parent recombination and volume recombination.
With the assumption that the recombination of e-ion pairs occurs with a 
probability of 50\%, from Eq.~\ref{eq:rec}, $\lambda = \frac{ln2}{t_{c}}$, 
gives the average recombination time of $t_{c} = 15$ ns ~\cite{Kubota}.
The final expression of the recombination probability can be expressed by Eq.~\ref{eq:rec3}:
\begin{equation}
r = 1-e^{-\frac{ln2}{t_{c}}(t_{pa} + t_{v})},
\label{eq:rec3}
\end{equation}
where $t_{pa}$ is the parent recombination time in the ionization zone and
$t_{v}$ is the volume recombination time. 
Note that the recombination begins as soon as some electrons
are thermalized, with an average thermalization time of 6.5 ns~\cite{sowa}, in the ionization zone. 
The parent recombination continues until all electrons become thermal electrons and the ionization zone 
expands into the plasma zone, the recombination is thus taken over by the volume recombination.   
Within the ionization zone, the Onsager radius, electron and its parent ion are very close together, they interact through
their Coulomb electric fields as isolated, individual particles. However, as the
distance between the electron and its parent ion increases beyond the Onsager radius, they interact
simultaneously with many nearby charged particles. This produces a collective interaction as a plasma. 
 
In the plasma zone,
the Coulomb force from any given charged particle causes all the nearby charges to move, thereby electrically polarizing the
medium. The charge particles are diffused through the ambipolar diffusion process, which is governed by the total number of charge
carriers and the strength of external field. 
Though the plasma time or the volume recombination time depends on the diffusion of charge carriers through multiple scattering,
a complicated process - it can be determined as a function of recoil energy using experimental data.
  
\subsection{Volume recombination time for non-zero field}
Since the volume recombination time depends on recoil energy under an electric field,
if one combines
Eq.~\ref{eq:number1} or Eq.~\ref{eq:number2} with Eq.~\ref{eq:rec}, it is seen that the volume recombination time
can be expressed in the form of a logarithm of recoil energy.
To fit the generic model (Eq.~\ref{eq:number1} or Eq.~\ref{eq:number2}) to data, the volume recombination time, $t_{v}$, 
equivalent to the plasma time, $t_{pl}$, is parametrized to be energy dependent:
\begin{equation}
t_{pl} = \alpha (\ln E_{r}) + \beta (\ln E_{r})^{2},
\label{eq:plas}
\end{equation}
where the quantities $\alpha$ and $\beta$ are electric field-dependent and recoil type dependent parameters, and can be parametrized by:
\begin{equation}
\alpha = \gamma_{1} F^{\delta_{1}},
\label{eq:plas1}
\end{equation}
and 
\begin{equation}
\beta = \gamma_{2} F^{\delta_{2}},
\label{eq:plas2}
\end{equation}
where F is the applied electric field, and $\gamma_{1}$, $\gamma_{2}$, $\delta_{1}$ and $\delta_{2}$ are free parameters. 

As introduced in Section 1, due to the weak field dependence in the charge yield of NRs, we assume $\alpha$ has negligible field dependence, which is determinable with a set of 
experimental data under a given field. However,   
$\beta$ is electric field dependent. In the case of ERs, because of a stronger field dependence in 
the charge yield of ERs, both $\alpha$ and $\beta$ are assumed to be field dependent. 

At non-zero field, the above parametrization functions are chosen based on the 
physical arguments that the plasma time depends on the initial ionization density, the diffusion of charge carriers, 
and the strength of the electric field. The energy-dependent terms arise from the density of the plasma track that confines the diffusion of thermal electrons 
in Coulomb electric field. 
When the plasma zone is formed on the track of the ionizing
particle, effective recombination of the thermal
electrons with the ions takes place in an ambipolar
diffusion process. As the recombination
proceeds, the ion density in the zone decreases, until ultimately a time is reached when the positive space
charge of the zone cannot retain the electrons. 
Ideally, two processes can cause diffusion and erosion of the plasma track: (1) the thermal electrons at the edge of plasma zone 
begin to drift away, under the influence of the external electric field and (2) the recombination within the plasma zone reduces the density of the plasma track. 
In realty, the plasma zone is a dynamic zone in which diffusion, recombination, and Coulomb
attraction alternate on thermal electrons. This dynamic process makes the accurate calculation of the plasma time difficult. 
However, the plasma time can be determined
with reliable experimental data. 

Using Eq.~\ref{eq:number1}, Eq.~\ref{eq:number2}, Eq.~\ref{eq:rec3}, and Eq.~\ref{eq:plas}, the charge and
light yield can be expressed as:
\begin{equation}
Q_{y} = \frac{L}{W_{i}}e^{\{-\frac{ln2}{t_c}[t_{pa}+\alpha(lnE_{r})+\beta(lnE_{r})^{2}]\}},
\label{eq:cy1}
\end{equation}
and 
\begin{equation}
L_{y} = \frac{L}{W_{i}}(1-e^{\{-\frac{ln2}{t_c}[t_{pa}+\alpha(lnE_{r})+\beta(lnE_{r})^{2}]\}}+\frac{N_{ex}}{N_{i}}),
\label{eq:ly1}
\end{equation}

\subsubsection{Parameters determination for NRs}

For NRs, the recently released LUX D-D neutron data~\cite{Akerib} 
show both charge and light yield
as a function of nuclear recoil energy. Since the charge and light yield are anti-correlated, 
one can use the LUX charge yield data
to determine the plasma time for NRs. If the model works, it should describe the light yield with an anti-correlation to the charge
yield from the LUX data. When using Eq.~\ref{eq:cy1} to fit the LUX D-D neutron data, $L$ is the Lindhard quenching factor described 
in Eq.~\ref{eq:lindh}~\cite{lind}, $W_{i}$ is described in Eq.~\ref{eq:wconst}~\cite{Taka,Stein,Roberts,Prunier}, 
$t_{c}$ = 15 ns, $\frac{N_{ex}}{N_{i}}$ is described in Eq.~\ref{eq:ratio}, $E_{r}$ is nuclear recoil
energy in keV.

\begin{equation}
L = \frac{k\cdot g(\epsilon)}{1+k\cdot g(\epsilon)},
\label{eq:lindh}
\end{equation}
where $k$ = 0.133$Z^{2/3} A^{-1/2}$,
$g(\epsilon) = 3\epsilon^{0.15} + 0.7\epsilon^{0.6} + \epsilon$, and
$\epsilon$ = 11.5$E_{r} Z^{-7/3}$ for a given atomic number, $Z$, mass number,
$A$, and recoil energy, $E_r$.
\begin{equation}
W_{i} = 14.94 + 8.35 \times \frac{N_{ex}}{N_{i}}.  
\label{eq:wconst}
\end{equation}

Since $\frac{N_{ex}}{N_{i}}$ represents the ratio of probability of direct excitation to ionization,
it can be parameterized as a function of electron-equivalent recoil energy using the ratio of the cross-section 
of excitation to ionization:
\begin{equation}
\frac{N_{ex}}{N_{i}} = \frac{1-\exp(-I/E_{er})}{3 + \exp(-I/E_{er})},
\label{eq:ratio}
\end{equation}
where $I$ is the mean ionization potential of xenon and $E_{er}$ is electron-equivalent recoil energy.
At very low energies, the term with a constant of 3 constrains $\frac{N_{ex}}{N_{i}}$ to be a maximum value of $\sim$ 30\%, 
 in accordance with the acceptable range of 6\% to 26\% by~\cite{Taka, NEST, MMI, MSZY, Doke, Dahl, Kubota}.
The experimental values for $I$ from Barkas and Berger~\cite{Exitationpotential, barkas}
can be used to determine $I$ = 555.57 eV for xenon.

\begin{figure}[htb!!]
\begin{center}
\includegraphics[angle=0,width=12.cm] {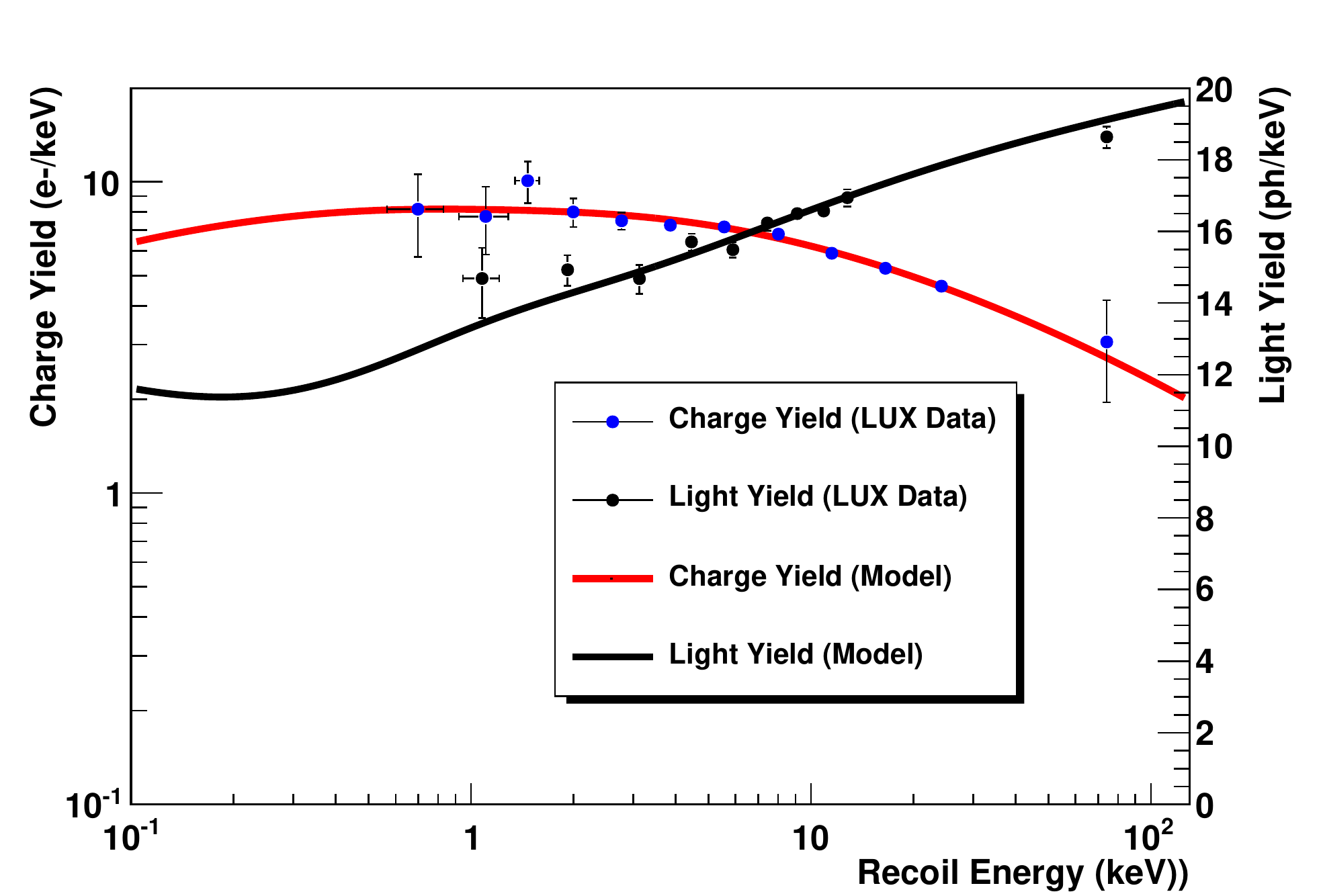}
\caption{\small {Determination of $t_{pa}$, $\alpha$ and $\beta$ by fitting Eq.~\ref{eq:cy1} and Eq.~\ref{eq:ly1} into the LUX D-D neutron data.
$t_{pa}$ = 1.5$\pm$0.5 ns, $\alpha$ = 3.617$\pm$0.521 ns  and $\beta$ = 1.313$\pm$0.196 ns  at 181 V/cm with $\chi^{2}/ndf$ = 4.299/9. The word ``Model'' in the legend represents the generic model with the Wang-Mei's recombination probability from this work. }}
\label{fig:luxdd}
\end{center}
\end{figure}

Figure~\ref{fig:luxdd} illustrates the determination of the parameters, $t_{pa}$, $\alpha$ and $\beta$, using the LUX D-D
neutron data. As expected, once the values of $t_{pa}$, $\alpha$ and $\beta$ are determined using the charge yield from the LUX
D-D neutron data, the model also describes the light yield data very well as shown in Figure~\ref{fig:luxdd}.

\begin{figure}[htb!!]
\begin{center}
\includegraphics[angle=0,width=12.cm] {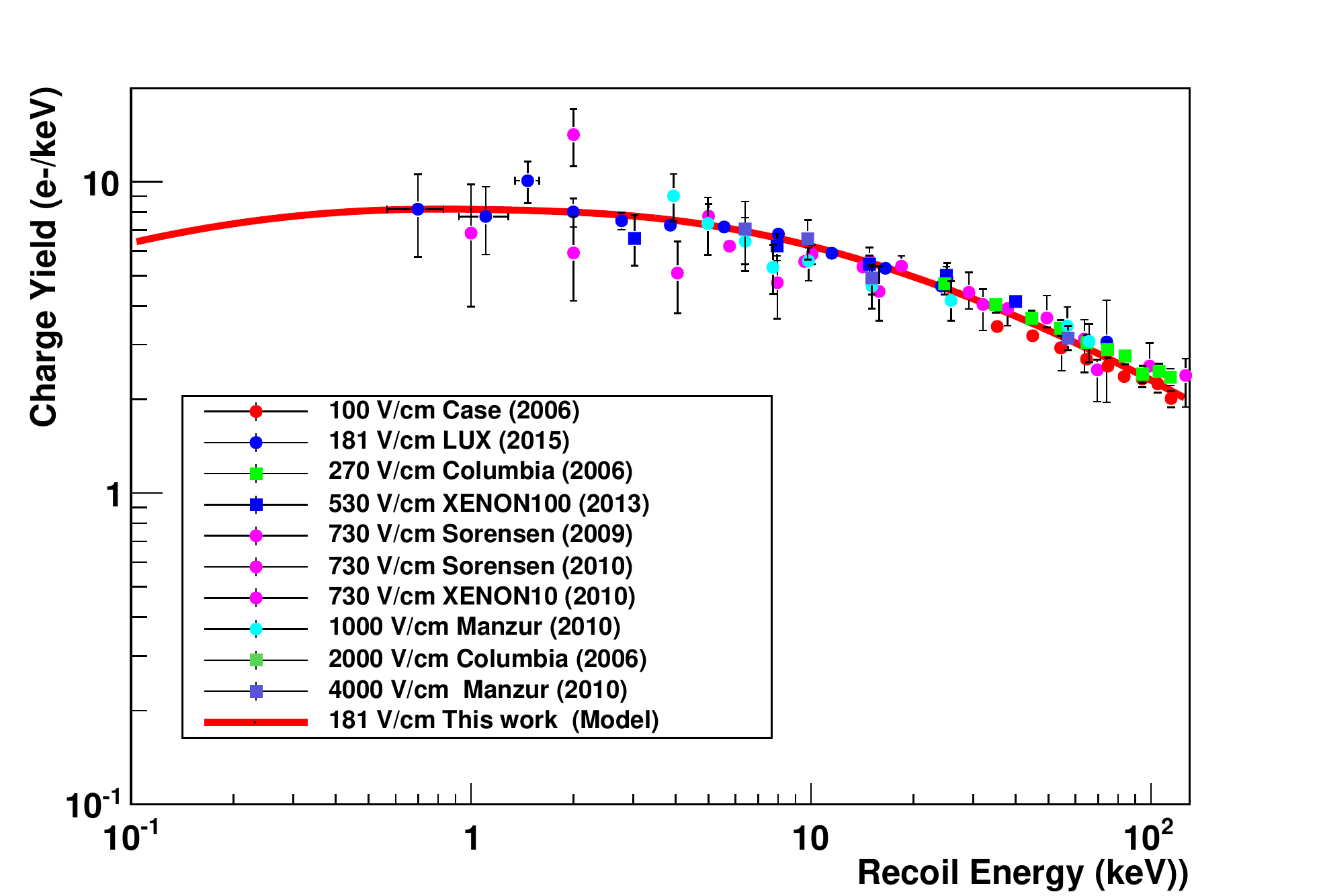}
\caption{\small {A charge yield comparison between the generic model in Eq.~\ref{eq:cy1}  
and data taken with different field strengths using the plasma time
determined from the LUX data. The data in this plot are from Case (2006) at 100 V/cm~\cite{case}, 
LUX D-D (2015) at 181 V/cm~\cite{Akerib}, 
Columbia (2006) at 270 V/cm~\cite{case}, XENON100 (2013) at 530 V/cm~\cite{xe100ap}, Sorensen (2009) at 730 V/cm~\cite{sor1},
Sorensen (2010) at 730 V/cm~\cite{sor2}, XENON10 (2010) at 730 V/cm~\cite{xe10e}, Manzur (2010) at 1000 V/cm~\cite{manz}, 
Columbia (2006) at 2000V/cm~\cite{case}, and Manzur (2010) at 4000 V/cm~\cite{manz}. 
The word ``Model'' in the legend represents the generic  model with the Wang-Mei's recombination probability from this work.}}
\label{fig:alldatacharge}
\end{center}
\end{figure}

One can fit the model, Eq.~\ref{eq:cy1}, into the data available from experiments with 
different electric
fields to look into the field dependence of $t_{pa}$, $\alpha$ and $\beta$. 
Before doing so, we notice that $t_{pa}$ is independent of electric field since
the external field cannot penetrate the ionization zone. Therefore, we treat
$t_{pa}$ as constant. 
Figure~\ref{fig:alldatacharge} shows a weak dependence on the electric field. 

Furthermore,  we found that $\alpha$ is independent of the field and the value of $\delta_{1}$ = 0 in Eq.~\ref{eq:plas1} and $\gamma_{1}$ = $\alpha$.
The values of $\beta$ is found to be weakly depended on
the field and  the value of $\delta_{2}$ in Eq.~\ref{eq:plas2} is found by the best fit to various $\beta$ values
in Table~\ref{table:beta} with respect to different electric fields
and can be interpreted as a weak field-dependence in the volume recombination. 

\subsubsection{Parameters determination for ERs}

\begin{table}[htb!!]
\caption{The obtained values of $\beta$ in $t_{pl}$ for NRs with respect to different fields with a fixed $\alpha$ = 3.617$\pm$0.521 ns.}
\label{table:beta}
\begin{center} 
\begin{tabular}{|c|c|c|}
\hline
F & $\beta$ & $\chi^{2}$/ndf\\
(V/cm) & (ns) &\\
\hline
100  & 1.409$\pm$0.199 &9.335/8\\       
\hline 
181  & 1.313$\pm$0.196&4.299/9 \\
\hline
270  & 1.243$\pm$0.223 &1.408/9\\
\hline
530  & 1.153$\pm$0.148 &3.274/4\\
\hline
730  & 1.129$\pm$0.132 &32.25/22\\
\hline
1000  & 1.096$\pm$0.159 &7.149/8\\
\hline
2030  & 1.025$\pm$0.120 &4.951/9\\
\hline
4000  & 0.962$\pm$0.107 &1.269/3\\
\hline
\end{tabular}
\end{center}
\end{table}

\begin{figure}[htb!!]
\begin{center}
\includegraphics[angle=0,width=12.cm] {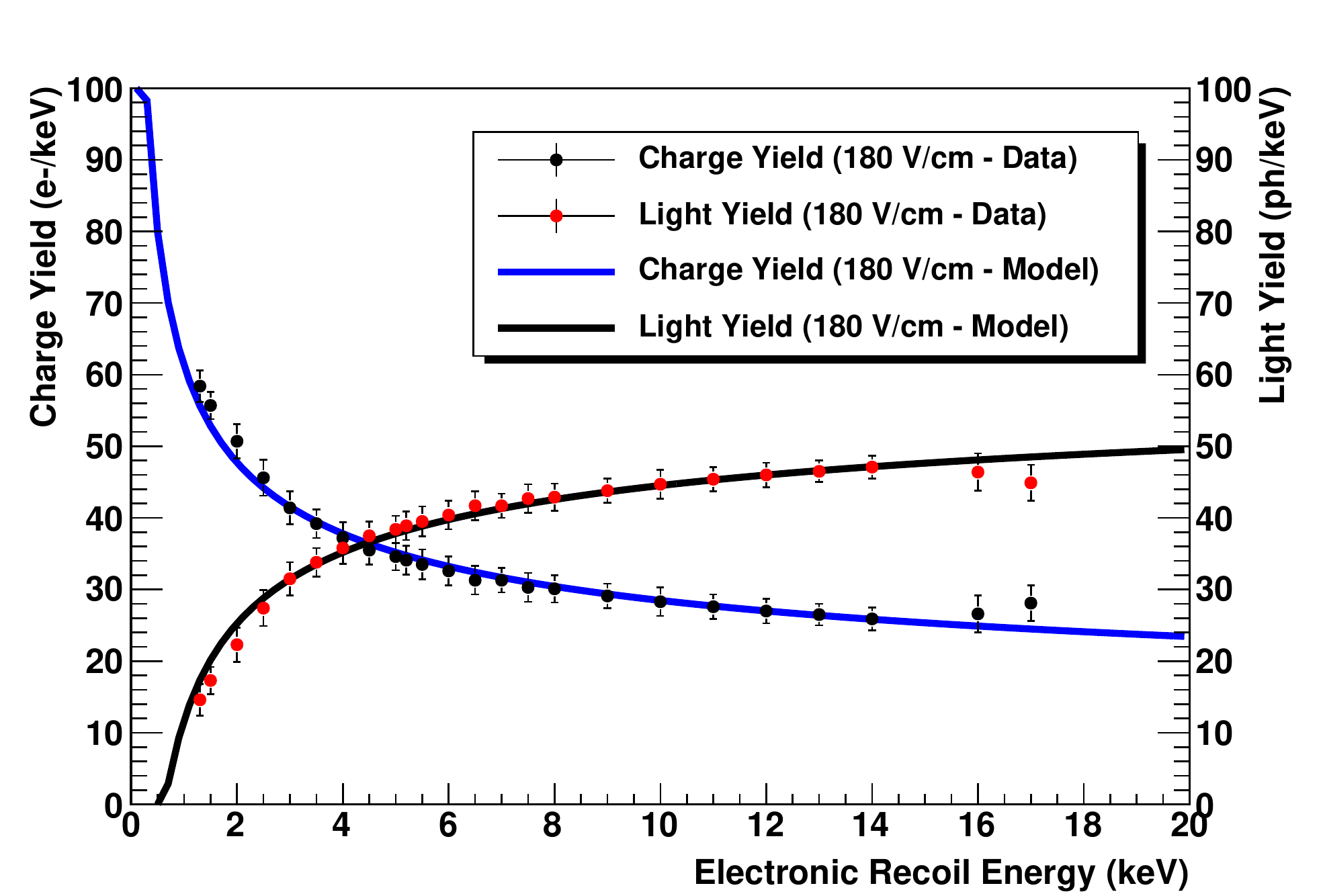}
\caption{\small {Determination of $t_{pa}$, $\alpha$ and $\beta$ by fitting Eq.~\ref{eq:cy1} and Eq.~\ref{eq:ly1} into the LUX tritium data with 180 V/cm~\cite{luxtr}. The model predicts that
$t_{pa}$ = 1.5$\pm$0.3 ns, $\alpha$ = 7.425$\pm$0.147 ns and $\beta$ = -0.198$\pm$0.069 ns  at 180 V/cm with $\chi^{2}/ndf = 12.97/23$. The word ``Model'' in the legend represents the generic  model with the Wang-Mei's recombination probability from this work. }}
\label{fig:ertr}
\end{center}
\end{figure}

Similarly for ERs, one can use the LUX tritium data~\cite{luxtr} to determine $t_{pa}$, $\alpha$ and $\beta$ for ERs using Eq.~\ref{eq:cy1} and Eq.~\ref{eq:ly1} 
at which
$L$ is Lindhard quenching (Eq.~\ref{eq:lindh}), $W_{i}$ = 15.6 eV, and $\frac{N_{ex}}{N_{i}}$ = 0.1387. 
Figure~\ref{fig:ertr}
shows the best fit to the LUX tritium data obtained with an electric field of 180 V/cm. 

\begin{figure}[htb!!]
\begin{center}
\includegraphics[angle=0,width=12.cm] {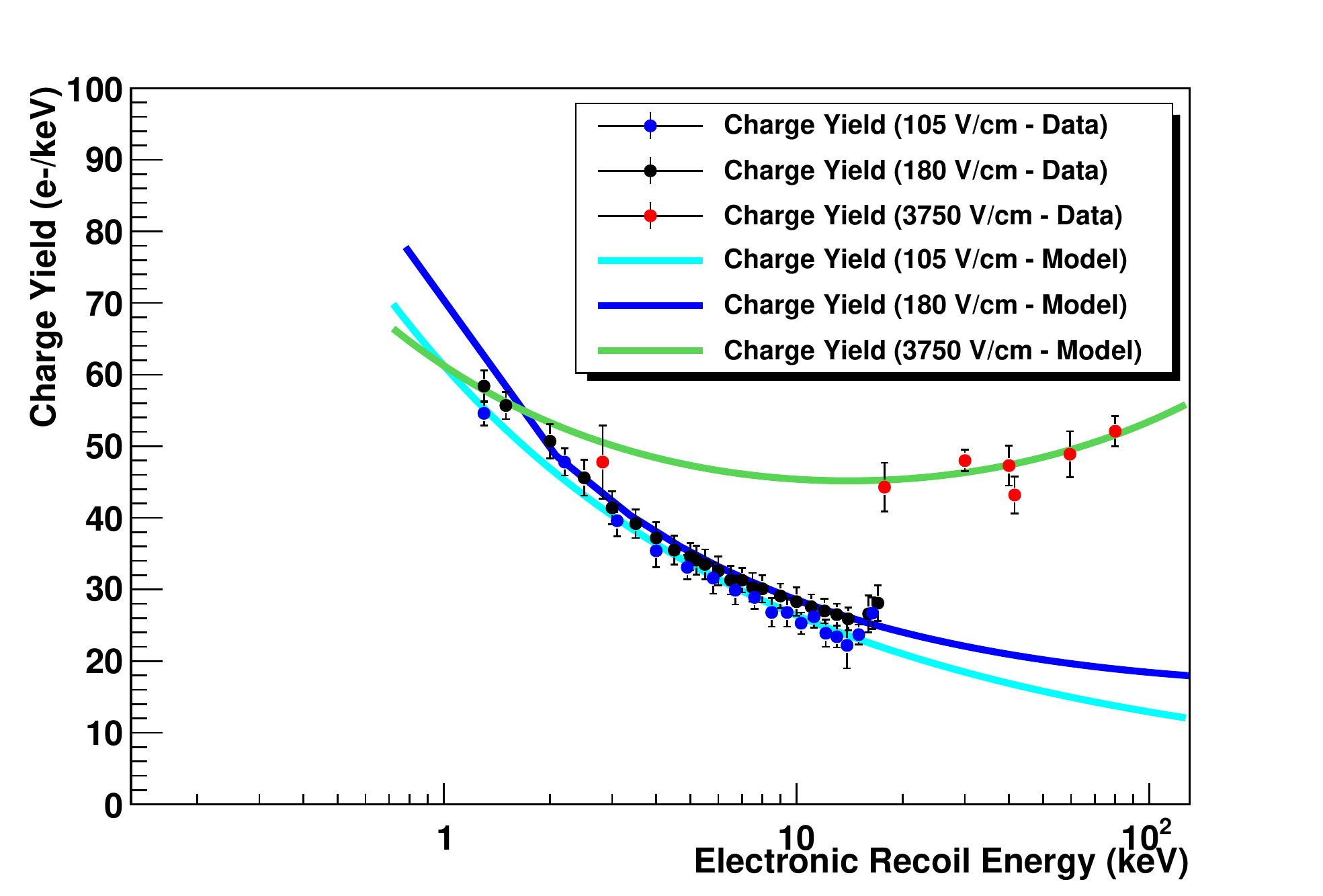}
\caption{\small {A charge yield comparison between the generic model in Eq.~\ref{eq:cy1}  
and data taken with different field strengths using the plasma time
determined from the LUX tritium data.The data is taken from 
the LUX tritium calibration obtained with 105 V/cm, 180 V/cm, and 
D. Yu. Akimov et al. (2014)~\cite{dyu}. The word ``Model'' in the legend represents the generic model with the Wang-Mei's recombination probability from this work.
}}
\label{fig:erdep}
\end{center}
\end{figure}

As stated above, the external field cannot penetrate the ionization zone, $t_{pa}$ is independent of electric field for ERs as well. 
The power-law field dependence of $\alpha$ and  $\beta$ is determined with two data sets taken from the LUX tritium calibration
at 105 V/cm and 180 V/cm as well as one data set at 3750 V/cm taken by D. Yu. Akimov et al. in 2014. 
Figure~\ref{fig:erdep} shows a weak dependence on electric field for data taken under relatively comparable electric fields. 
If the electric field varies dramatically,
the model defined by the parameters obtained under low electric fields (105 V/cm and 180 V/cm) 
 cannot fit well with data taken under a very high electric field (3750 V/cm).   

\begin{table}[htb!!]
\caption{The obtained values of $\alpha$ and $\beta$ for ERs with respect to different fields.}
\label{table:beta1}
\begin{center} 
\begin{tabular}{|c|c|c|c|}
\hline
F & $\alpha$ & $\beta$ &$\chi^{2}$/ndf\\
(V/cm) &  (ns ) &(ns) &\\
\hline
105  & 8.073$\pm$0.819 & -0.176$\pm$0.071 &8.467/15\\
\hline
180 & 7.425$\pm$0.147  & -0.198$\pm$0.069 &12.97/23\\      
\hline 
3750  & 4.678$\pm$0.927 & -0.899$\pm$0.240 &4.414/5\\
\hline
\end{tabular}
\end{center}
\end{table} 

\begin{figure}[htb!!]
\begin{center}
\includegraphics[angle=0,width=12.cm] {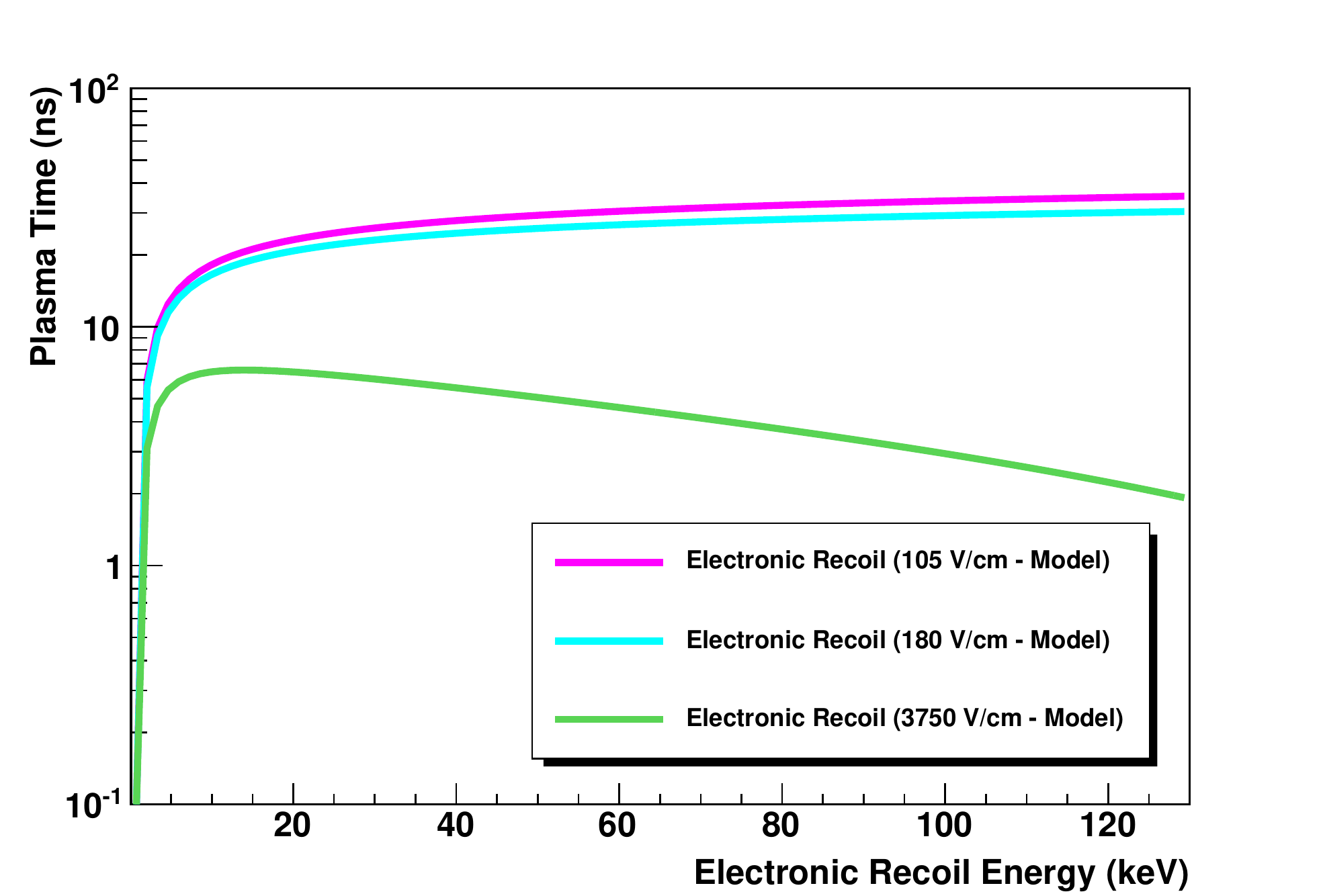}
\caption{\small{Shown is the plasma time, $t_{pl}$, for ERs under different electric fields. The word ``Model'' in the legend represents the generic model with the Wang-Mei's recombination probability from this work. }}
\label{fig:plasField}
\end{center}
\end{figure}

Once again, the field dependent values of $\delta_{1}$ and $\delta_{2}$ in Eq.\ref{eq:plas1} and Eq.~\ref{eq:plas2}
for ERs are found by the best fit to various $\alpha$ and $\beta$ values
in Table~\ref{table:beta1} with respect to the three electric fields, 
and can be interpreted as a stronger field dependence in the volume recombination.
Figure~\ref{fig:plasField} shows the field dependence of the plasma time for ERs. $t_{pl}$ is much larger at a lower electric field, 
which indicates more electrons recombine with ions in the volume recombination zone in contrast to that under a higher electric field.     

\begin{figure}[htb!!]
\begin{center}
\includegraphics[angle=0,width=12.cm] {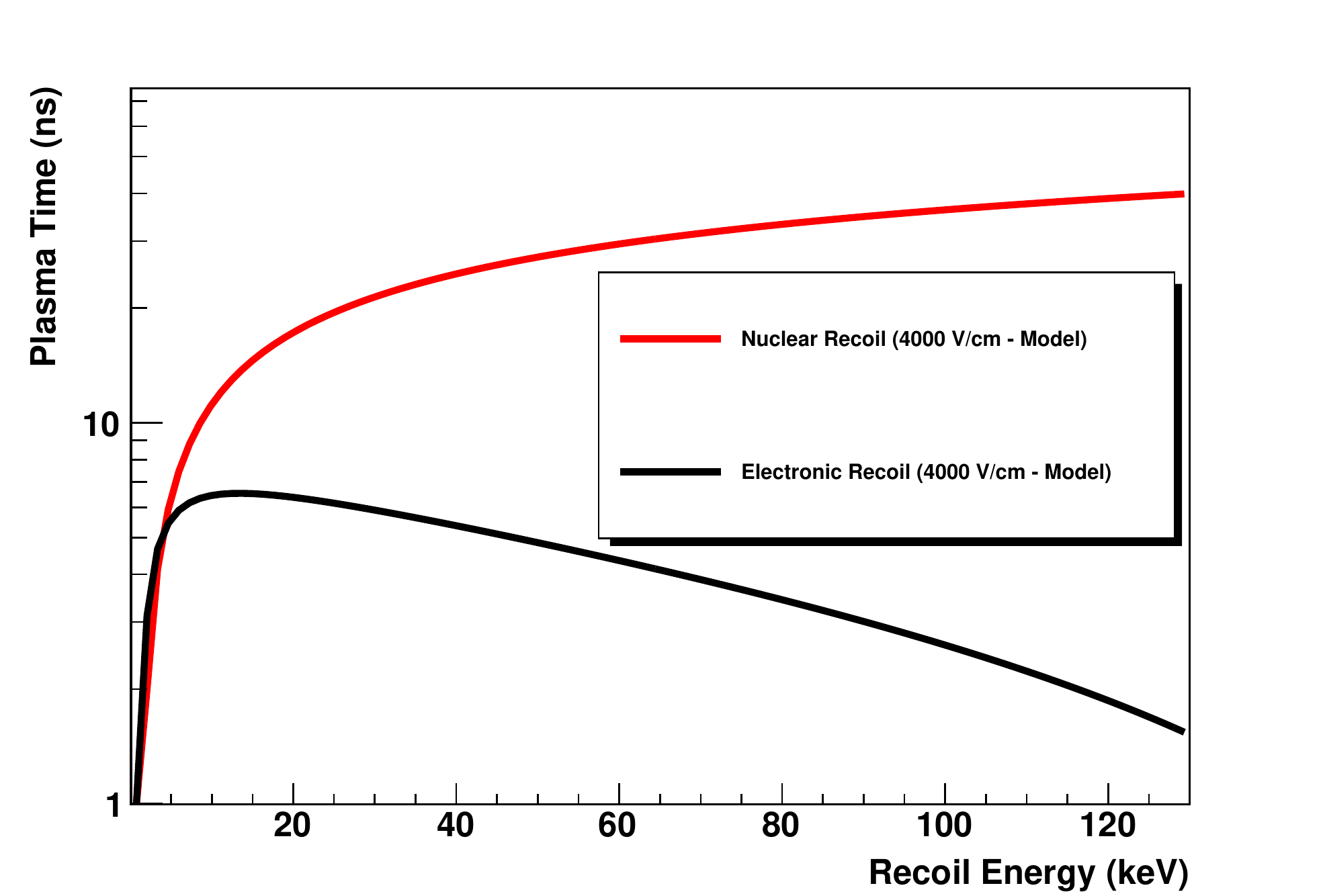}
\caption{\small{Shown is the plasma time, $t_{pl}$, as a function of recoil energy under an electric field of 4000 V/cm. The word ``Model'' in the legend represents the generic model with the Wang-Mei's recombination probability from this work.
}}
\label{fig:plastime}
\end{center}
\end{figure}

Figure~\ref{fig:plastime} shows the plasma time as a function of recoil energy for two types of events under the same electric field.
The plasma time for NRs is much larger than that for ERs at a field of 4000 V/cm. This is because
the plasma effect depends largely on the ionization density.
The higher density of ionization track for NRs induces a stronger internal electric field that shields 
the influence of external electric field so that the plasma zone can retain thermal electrons for a longer plasma time,
which results in more electrons recombining with ions in a higher density plasma volume. 
\begin{figure}[htb!!]
\begin{center}
\includegraphics[angle=0,width=12.cm] {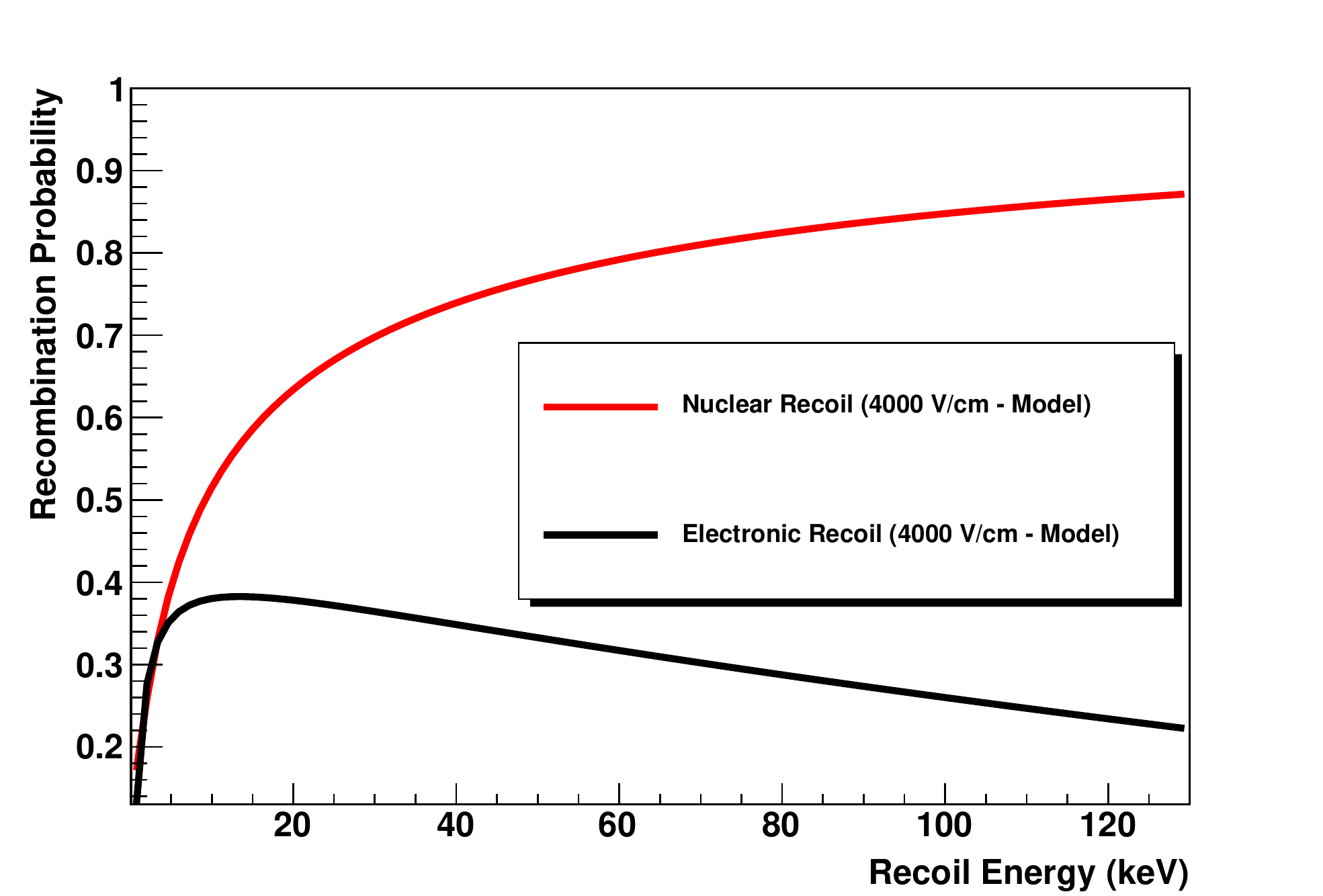}
\caption{\small{ Shown is the recombination probability  as a function of recoil energy at a non-zero field
with F = 4000 V/cm. The word ``Model'' in the legend represents the generic model with the Wang-Mei's recombination probability from this work.}}
\label{fig:compareRecnon}
\end{center}
\end{figure}
 
Combining the plasma time with the effect of the parent recombination time,  
Figure~\ref{fig:compareRecnon} shows the recombination probability as a function of recoil energy at 4000 V/cm,  
more electrons will recombine with ions for NRs than ERs at a given energy.  

We summarize the values of the parameters from Eq.~\ref{eq:plas1} and Eq.~\ref{eq:plas2} for NRs in Table~\ref{table:tab1}, and for ERs in Table~\ref{table:tab2}.
The uncertainty of these parameters for ERs is as large as $\sim$10\% due to the insufficient data at different fields
for the best fit. Therefore, more data taken under different fields would improve the accuracy of these parameters. 
\begin{table}[htb!!]
\caption{The obtained parameters for NRs in Eq.~\ref{eq:plas1} and Eq.~\ref{eq:plas2} with a $\chi^{2}$/ndf = 0.4407/6.}
\label{table:tab1}
\begin{center} 
\begin{tabular}{|c|c|}
\hline
$\alpha$ & 3.617$\pm$0.521\\
  ns &          \\
\hline 
$\gamma_{2}$ & 2.102$\pm$0.876  \\
  ns   &  \\
\hline
$\delta_{2}$ & -0.0943$\pm$0.0678  \\
\hline
\end{tabular}
\end{center}
\end{table}

\begin{table}[htb!!]
\caption{The obtained parameters for ERs in Eq.~\ref{eq:plas1} and Eq.~\ref{eq:plas2}. The value of 
$\chi^{2}$/ndf for obtaining $\gamma_{1}$ and $\delta_{1}$ is 0.0002307/1 and the value of $\chi^{2}$/ndf for 
obtaining $\gamma_{2}$ and $\delta_{2}$ is 0.06632/1.}
\label{table:tab2}
\begin{center} 
\begin{tabular}{|c|c|}
\hline
$\gamma_{1}$ & 16.39$\pm$5.31  \\
  ns &          \\
\hline
$\delta_{1}$ &  -0.1525$\pm$0.0622 \\
\hline
$\gamma_{2}$ & -0.01735$\pm$0.01393  \\
  ns  &  \\
\hline
$\delta_{2}$ & 0.4791$\pm$0.118 \\
\hline
\end{tabular}
\end{center}
\end{table}

\subsection{Volume recombination time for zero field}
Without an external electric field, 
the volume recombination time
$t_{v}$ depends on the initial number of e-ion pairs and ambipolar diffusion,
and is expected to be energy dependent with the same parametrization function as Eq.~\ref{eq:plas}.
The free parameters $\alpha$ and $\beta$ are obtained by normalizing the scintillation efficiency from the
generic model (Eq.\ref{eq:ly1})
to the available data for ERs~\cite{eap, Baudis},
and NRs~\cite{manz,arne,akim,apri,plan,chep,apri9}. The functions describing the volume recombination time are found to be:

ERs: 
\begin{equation}
t_{v}^{er} = (4.15\pm0.12) \times 10^{3} \times (\ln E_{er}) + (3.444\pm0.101)\times (\ln E_{er})^{2}, 
\label{eq:plaser}
\end{equation}   
where the fitted $\chi^{2}$/ndf = 1.234/4.

NRs:
\begin{equation}
t_{v}^{nr} = (14.712\pm1.14)\times (\ln E_{nr}) + (2.444\pm0.132) \times (\ln E_{nr})^{2},
\label{eq:plasnr}
\end{equation} 
where the fitted $\chi^{2}$/ndf = 1.726/5. 

\begin{figure}[htb!!]
\begin{center}
\includegraphics[angle=0,width=12.cm] {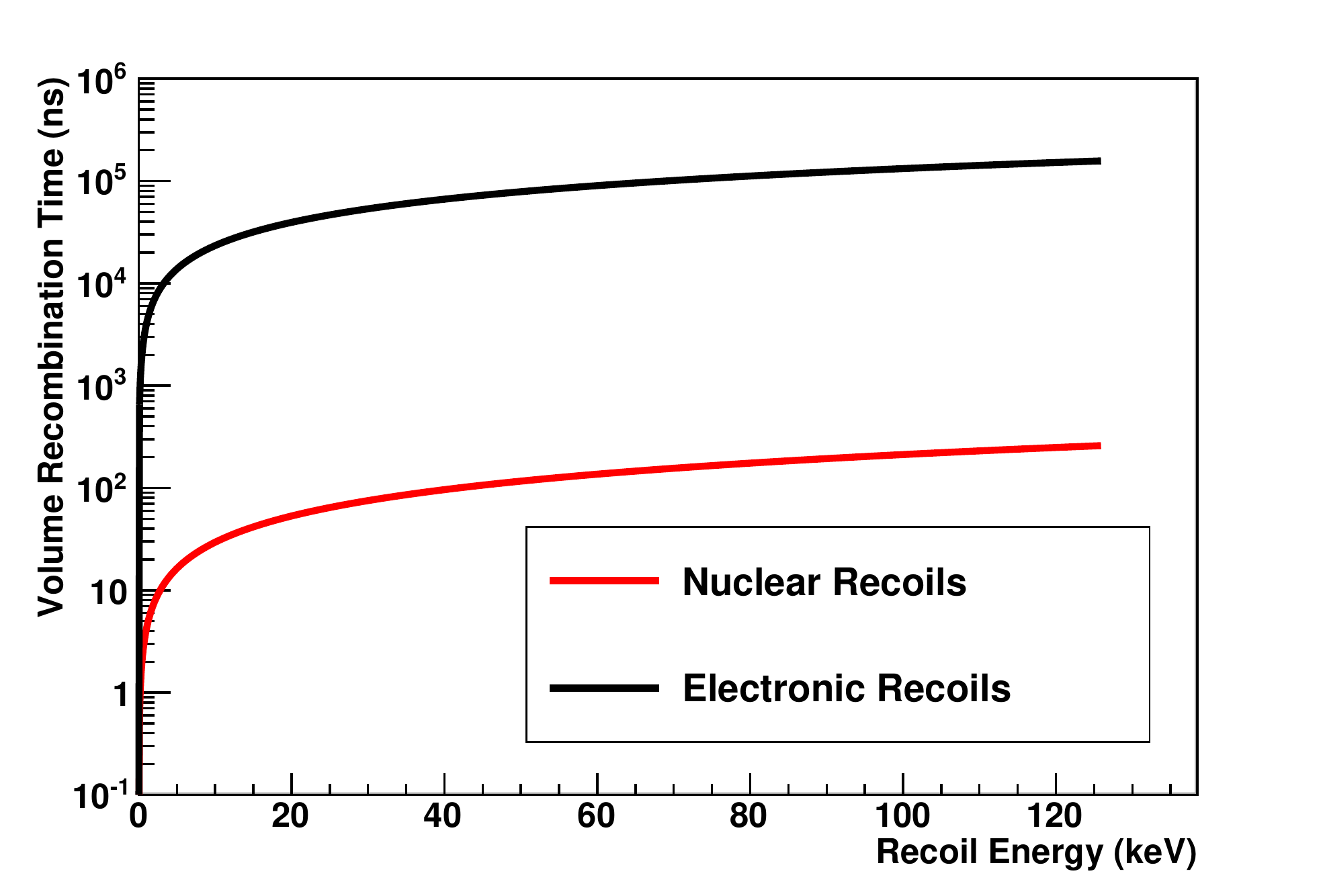}
\caption{\small{ Shown is the volume recombination time at zero field as a function of recoil energy. The
curves are obtained using the
generic model with the Wang-Mei's recombination probability from this work.}}
\label{fig:volreczero}
\end{center}
\end{figure}

\begin{figure}[htb!!]
\begin{center}
\includegraphics[angle=0,width=12.cm] {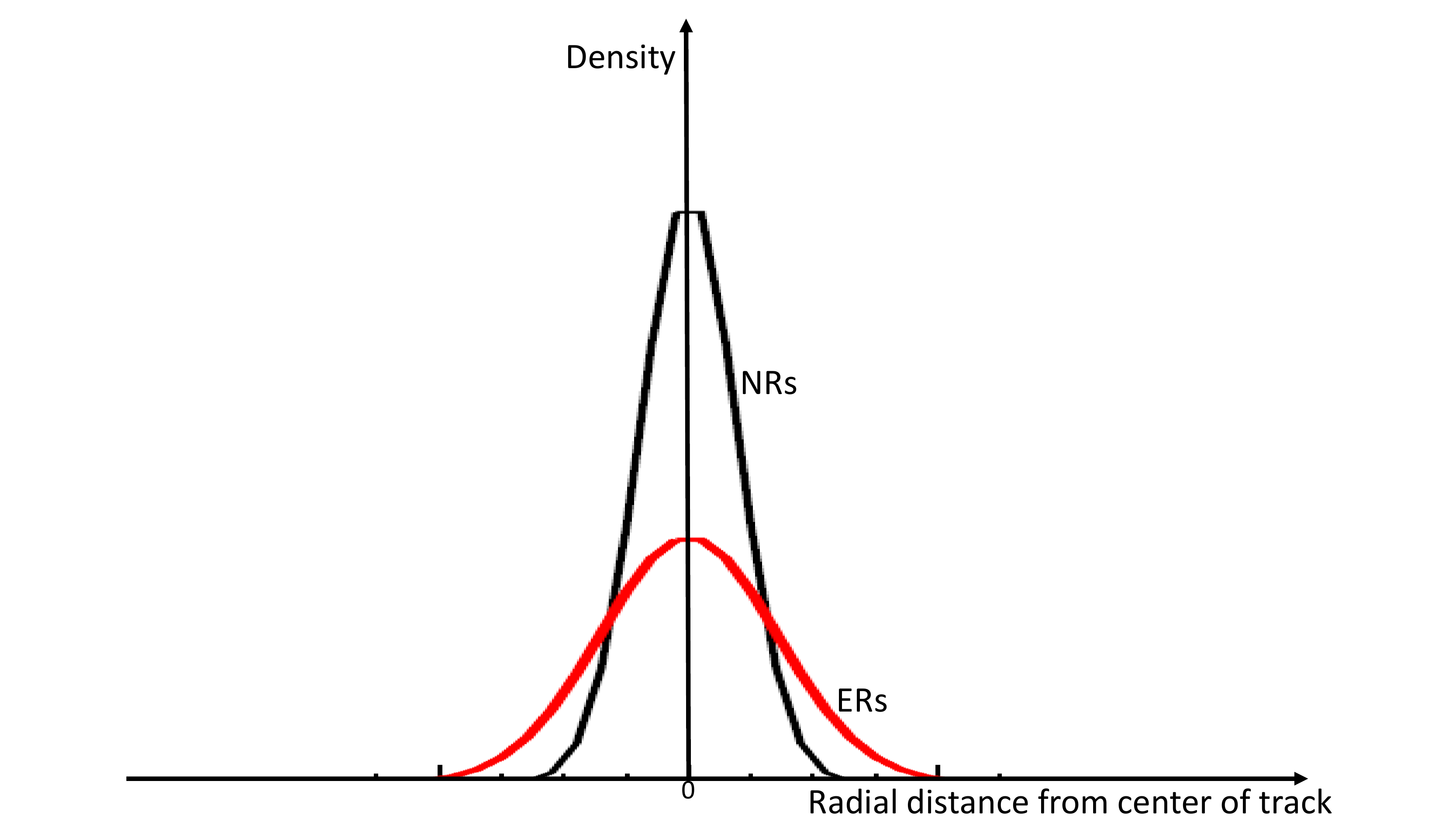}
\caption{\small{ Shown is a sketch of the density of plasma tracks versus the distance from center of 
the plasma tracks for ERs and NRs.}}
\label{fig:plasmatrack}
\end{center}
\end{figure}

As shown in Figure~\ref{fig:volreczero},
$t_{v}$ of NRs is much smaller than that of ERs. This is because 
the ionization density of ERs is relatively smaller than that of NRs. 
A smaller ionization density for ERs generates a weaker internal electric field,
allowing the electrons to diffuse farther away from the ionization track
as illustrated in Figure~\ref{fig:plasmatrack}.
Without an electric field, these electrons can undergo multiple scattering 
before recombining with ions.  

\begin{figure}[htb!!]
\begin{center}
\includegraphics[angle=0,width=12.cm] {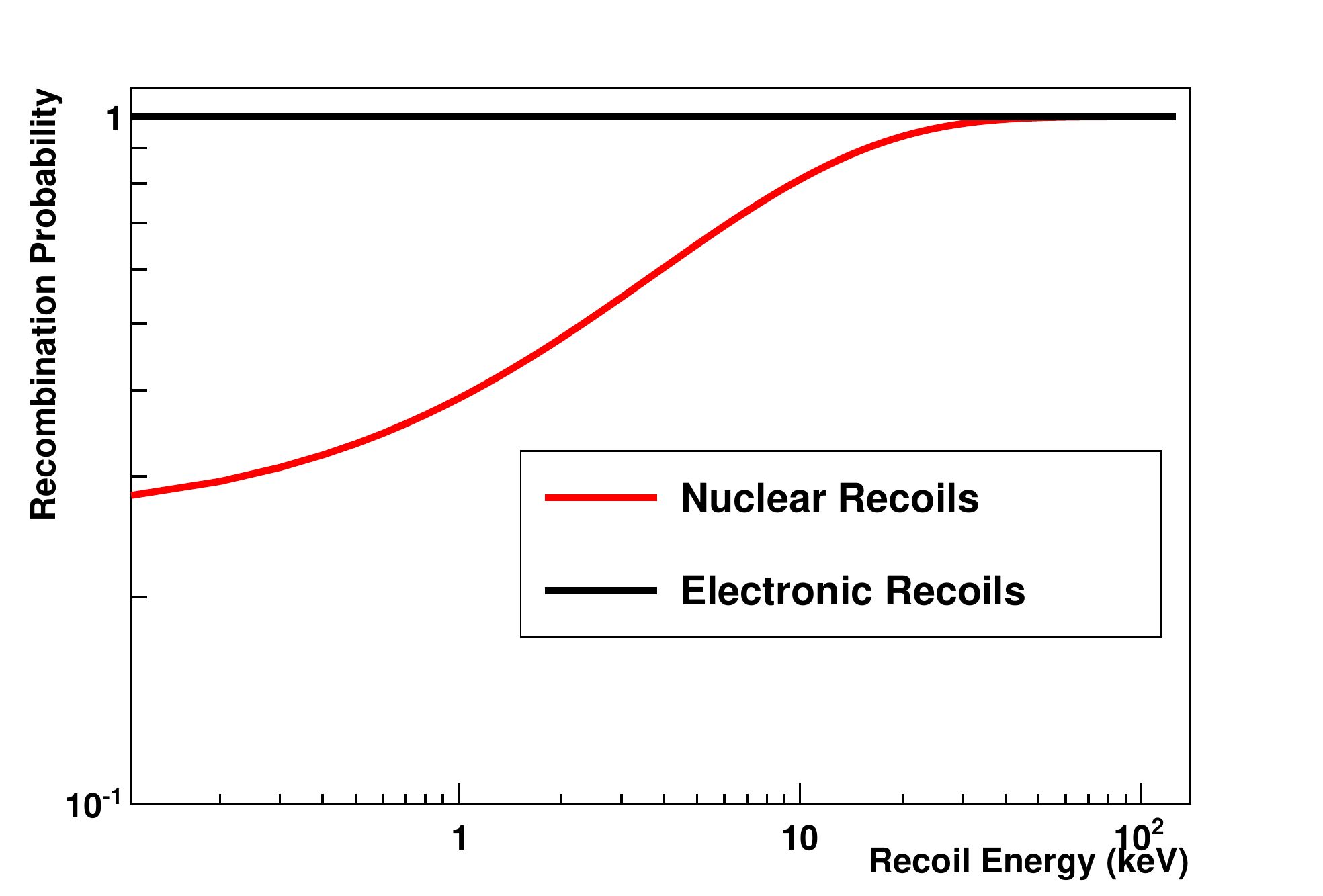}
\caption{\small{ Shown is the recombination probability as a function of recoil energy at zero field. The curves
are obtained using the generic model with the Wang-Mei's recombination probability from this work.}}
\label{fig:compareRec}
\end{center}
\end{figure}

Figure~\ref{fig:volreczero} 
shows the values of $t_{v}$ for
 ERs have a good agreement with ~\cite{Dolg}.
 At zero field, for both NRs and ERs, $t_{v} > t_{pa}$: 
more electrons recombine with ions within the volume recombination zone. 
As a result, shown in Figure~\ref{fig:compareRec}, the recombination probability for ERs is 1; for NRs, it is slightly dependent on recoil energy
in the region below 30 keV. This is because the ionization tracks for NRs with energy below 30 keV is shorter than Debye length~\cite{debye} 
and cannot
form any significant plasma effect. Therefore, the electrons are diffused away from the ionization track.
 Hence, the recombination probability
decreases as the recoil energy decreases. 

Utilizing the recombination model in this work, one can calculate the recombination probability at 
any given field and zero field for a liquid xenon detector. This recombination model is valid over
the energy range of 0.1 $-$ 130 keV.

\section{Implementation of the Model}
\label{Imple}

\subsection {Non-zero field}
\begin{figure}[htb!!]
\begin{center}
\includegraphics[angle=0,width=12.cm] {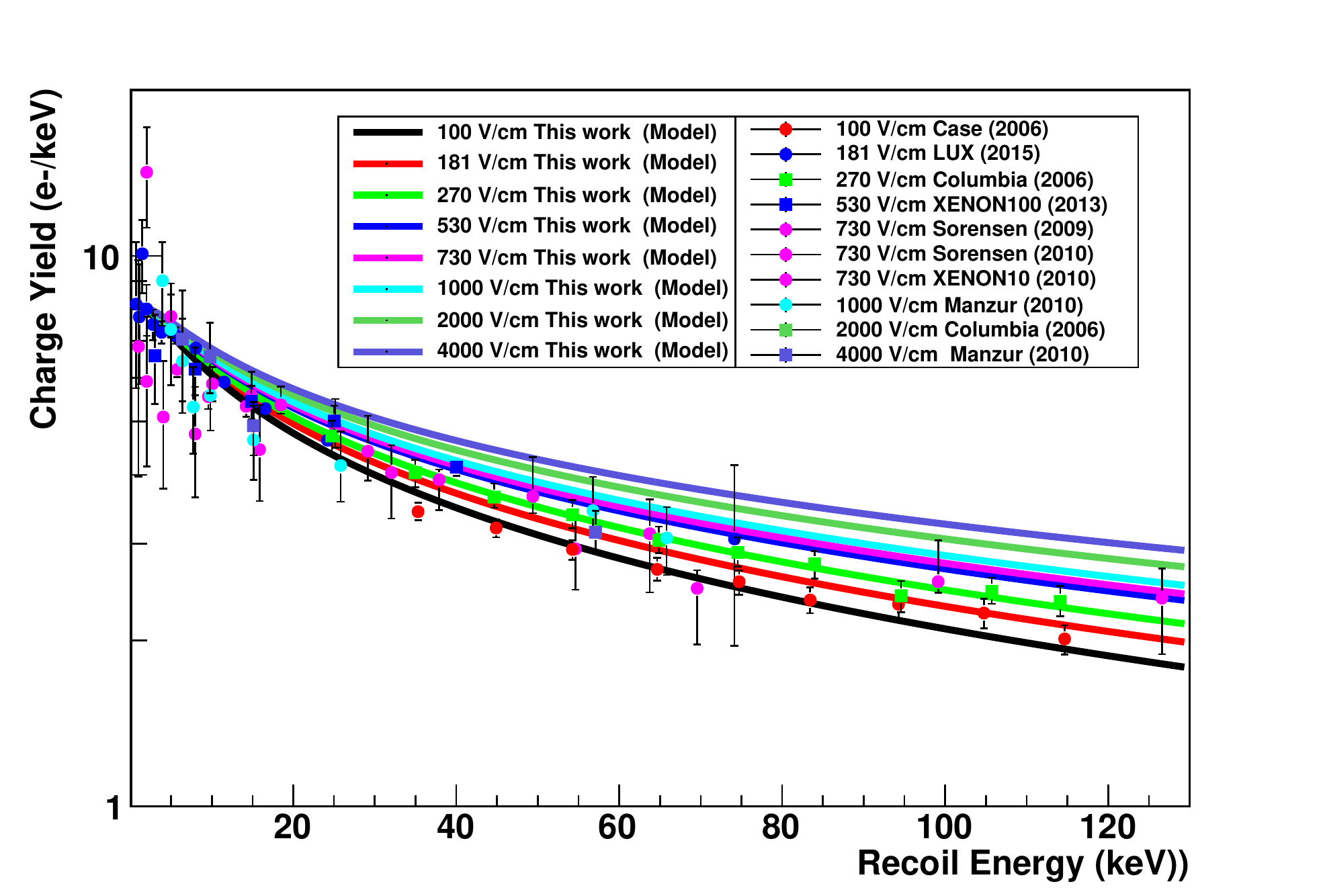}
\caption{\small{All measurements of the charge yield of NRs in comparison with the generic model in Eq.~\ref{eq:cy1} at different fields.
The data in this plot are from Case (2006) at 100 V/cm~\cite{case}, LUX D-D (2015) at 181 V/cm~\cite{Akerib}, 
Columbia (2006) at 270 V/cm~\cite{case}, XENON100 (2013) at 530 V/cm~\cite{xe100ap}, Sorensen (2009) at 730 V/cm~\cite{sor1},
Sorensen (2010) at 730 V/cm~\cite{sor2}, XENON10 (2010) at 730 V/cm~\cite{xe10e}, Manzur (2010) at 1000 V/cm~\cite{manz}, Columbia (2006) at 2000
V/cm~\cite{case}, and Manzur (2010) at 4000 V/cm~\cite{manz}. The word ``Model'' in the legend represents the generic model 
with the Wang-Mei's recombination probability from this work.
}}
\label{fig:NRchargenonzero}
\end{center}
\end{figure}

\begin{figure}[htb!!]
\begin{center}
\includegraphics[angle=0,width=12.cm] {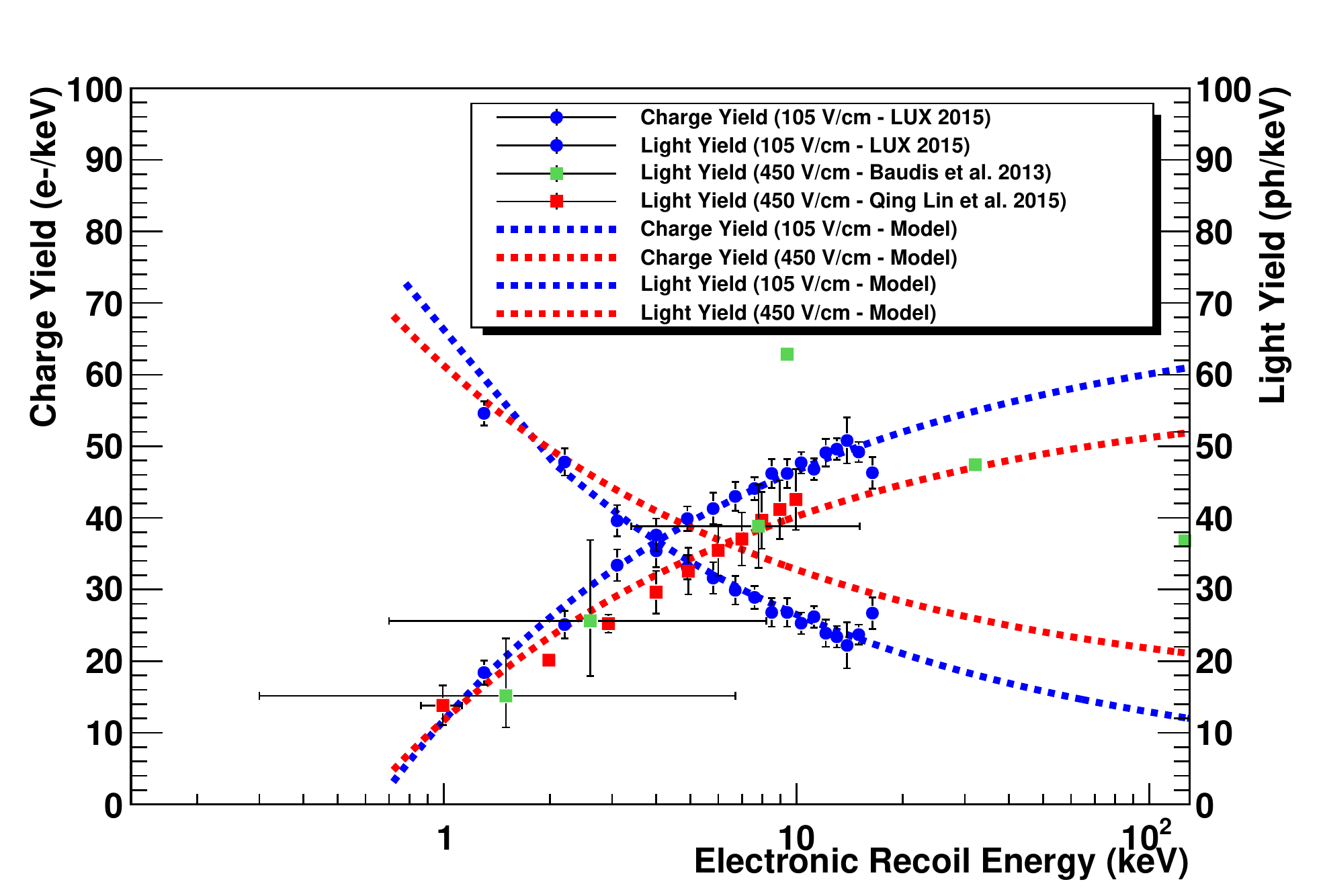}
\caption{\small{The charge yield of ERs in comparison with the generic model in Eq.~\ref{eq:cy1}. The data is taken from 
the LUX tritium calibration obtained with 105 V/cm, Qing Lin et al. (2015)~\cite{qing}, and Baudis et al. (2013). The word ``Model'' in the legend represents the generic model with the Wang-Mei's recombination probability from this work.
}}
\label{fig:ERchargenonzero}
\end{center}
\end{figure}

\begin{figure}[htb!!]
\begin{center}
\includegraphics[angle=0,width=12.cm] {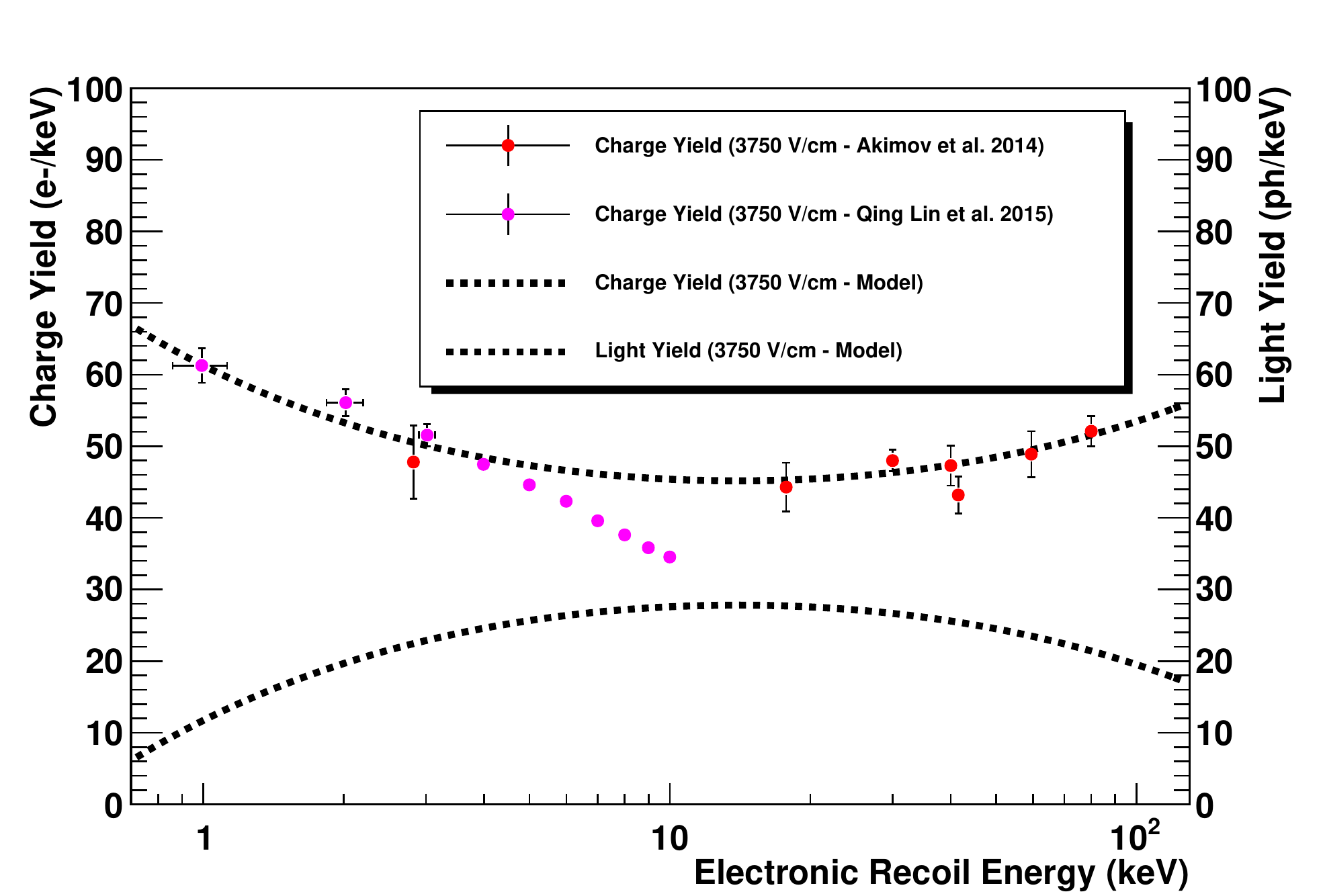}
\caption{\small{The charge yield of ERs in comparison with the generic model in Eq.~\ref{eq:cy1}. The data is taken from 
D. Yu. Akimov et al. (2014)~\cite{dyu} and Qing Lin et al.(2015)~\cite{qing}. The word ``Model'' in the legend represents the generic model with the Wang-Mei's recombination probability from this work.}}
\label{fig:ERchargenonzero1}
\end{center}
\end{figure}

At non-zero fields, the plasma time can be calculated with Eq.~\ref{eq:plas}, using the given values of $\alpha, \gamma$, and $\delta$ in Table~\ref{table:tab1} and Table~\ref{table:tab2}; 
the recombination probability 
is then obtained from Eq.~\ref{eq:rec3} for a given energy in the energy range of 0.1 $-$ 130 keV.  
The $W_{i}$-value is also energy dependent. 
Though the variation is small, its energy dependence could cause an uncertainty of 
5\% in energy reconstruction for recoil energy below 5 keV. Additionally, the Lindhard quenching factor is applied to 
the calculation of charge and light yield. 
The calculated charge yield from Eq.~\ref{eq:cy1} in units of $e^-/keV$ for NRs and ERs are compared
with available data in 
Figure~\ref{fig:NRchargenonzero}, Figure~\ref{fig:ERchargenonzero}, and Figure~\ref{fig:ERchargenonzero1}, respectively.
It is shown that the charge yield of ERs from this work (the model) agrees well with the data sets used in Figure~\ref{fig:ERchargenonzero}.
In Figure~\ref{fig:ERchargenonzero1}, our model agrees with the data from  D. Yu. Akimov et al. (2014)~\cite{dyu} and 
partly agree with the work from Qing Lin et al. (2015)~\cite{qing} based on the Thomas-Imel box model 
with the updated $4\xi$/$N_{i}$.

Clearly, for NRs, the model is weakly field dependent, and fits existing data very well. As discussed in Section 2.3.2, the plasma time for NRs
increases from a few nanoseconds to a few tens of nanoseconds. Hence, the charge yield shown in Figure~\ref{fig:NRchargenonzero} is energy dependent in the energy 
range of interest. 

However, for ERs, it is seen in Figure~\ref{fig:plasField}, the plasma time strongly depends
on the applied field. At a lower field, the plasma time ranges from a few nanoseconds to a few tens 
of nanoseconds. However, 
 the plasma time varies in a narrow range across the entire energy at a higher electric field 
($\sim$4000 V/cm),
the total recombination probability is also small as 
shown in Figure~\ref{fig:compareRecnon}. Therefore, the charge yield ($e^-/keV$) of ERs under a higher field is very weakly dependent on 
the recoil energy, shown in Figure~\ref{fig:ERchargenonzero}.
Whereas the volume recombination probability under low fields ($\sim$ a few hundreds V/cm) is much larger, 
due to the large plasma time that increases from a few nanoseconds to a few tens of nanoseconds 
as shown in Figure~\ref{fig:plasField}. 
Thus, the charge yield ($e^-/keV$) of ERs under a lower field is strongly dependent on the recoil energy.   

\subsection{Zero field}
\begin{figure}[htb!!]
\begin{center}
\includegraphics[angle=0,width=12.cm] {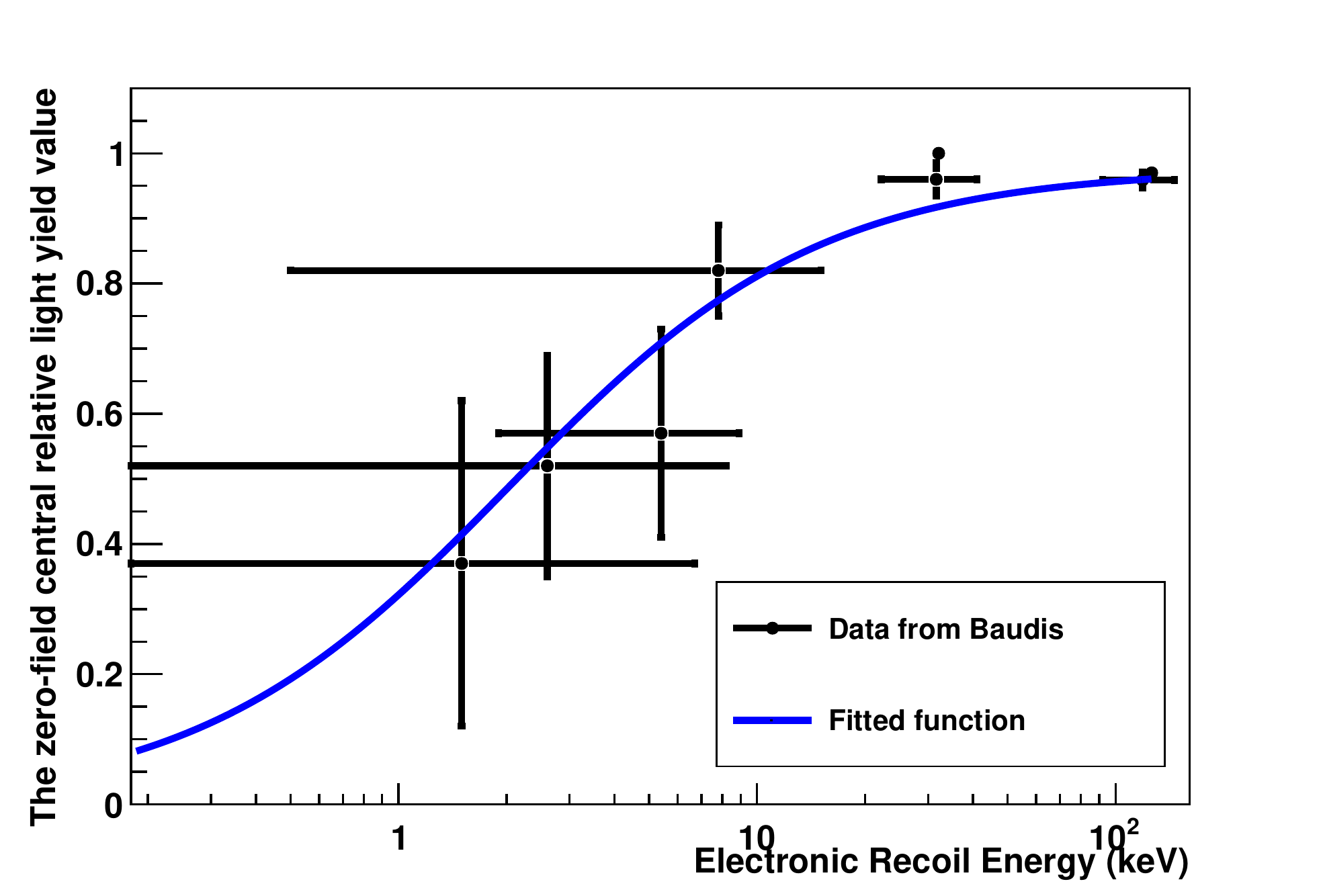}
\caption{\small{The zero-field central relative light yield value (relative to the scintillation emission at 32.1 keV) 
as a function of electronic recoil energy. The data points are from Baudis et al.~\cite{Baudis}. }}
\label{fig:erbau}
\end{center}
\end{figure}

\begin{figure}[htb!!]
\begin{center}
\includegraphics[angle=0,width=12.cm] {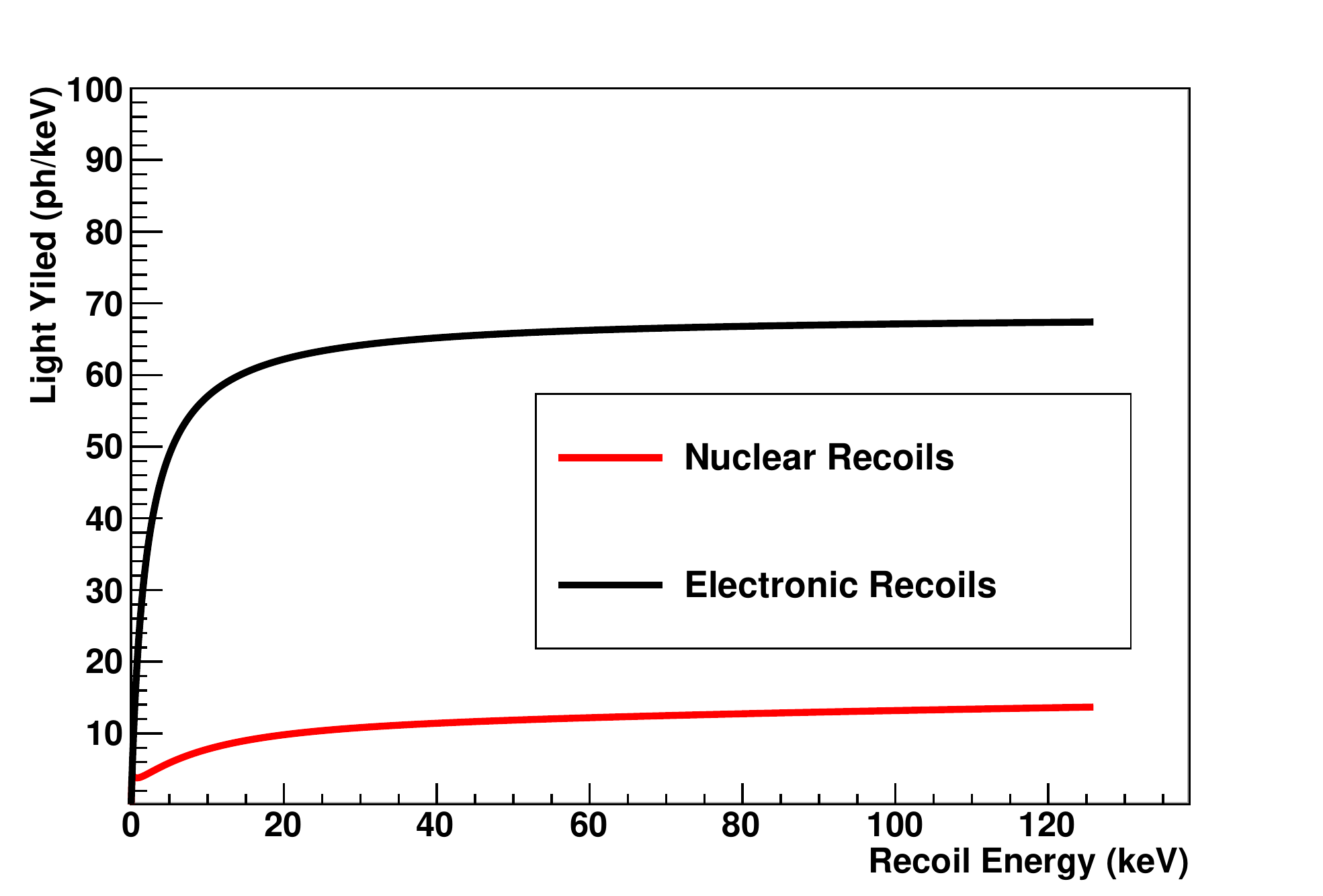}
\caption{\small{Light yield calculated from the generic model in Eq.~\ref{eq:number2} at zero field with the Wang-Mei's recombination probability from this work.}}
\label{fig:modelonly}
\end{center}
\end{figure}

At zero field, the volume recombination time is calculated from Eq.~\ref{eq:plaser} (Eq.~\ref{eq:plasnr}) for ERs (NRs), and 
then the recombination probability is found from Eq.~\ref{eq:rec3}.
The variation of the $W_{i}$-value, Eq.\ref{eq:wconst}, is applied. Both the Lindhard quenching factor (Eq.\ref{eq:lindh})
 and the Hitachi quenching factor, 0.68~\cite{ahp}, 
are applied in the light yield for NRs. The quenching factor 
$q_{e}$ (Eq.~\ref{eq:photon}), from the best-fit to the
data~\cite{Baudis} as shown in Figure~\ref{fig:erbau}, is applied to ERs.
The fitted function with a $\chi^{2}$/ndf = 3.55/3 following the form of Birks' law can be expressed as:
\begin{equation}
q_{e} = \frac{0.982\pm0.018E_{er}}{1.999\pm0.003+0.96\pm0.02E_{er}}.
\label{eq:photon}
\end{equation}   
Thus, the quenching factor, $L$ = $q_{e}$, in Eq.~\ref{eq:number2} for ERs.
Note that this is different to the explanation of the reduction of the light yield using
 the existence of the effect of escaping electrons in the absence of an electric field~\cite{Doke}. 
The resulting light yield at zero field for ERs and NRs, shown in Figure~\ref{fig:modelonly}, indicates 
there exists a light reduction for ERs and NRs at very low energies. 

\begin{figure}[htb!!]
\begin{center}
\includegraphics[angle=0,width=12.cm] {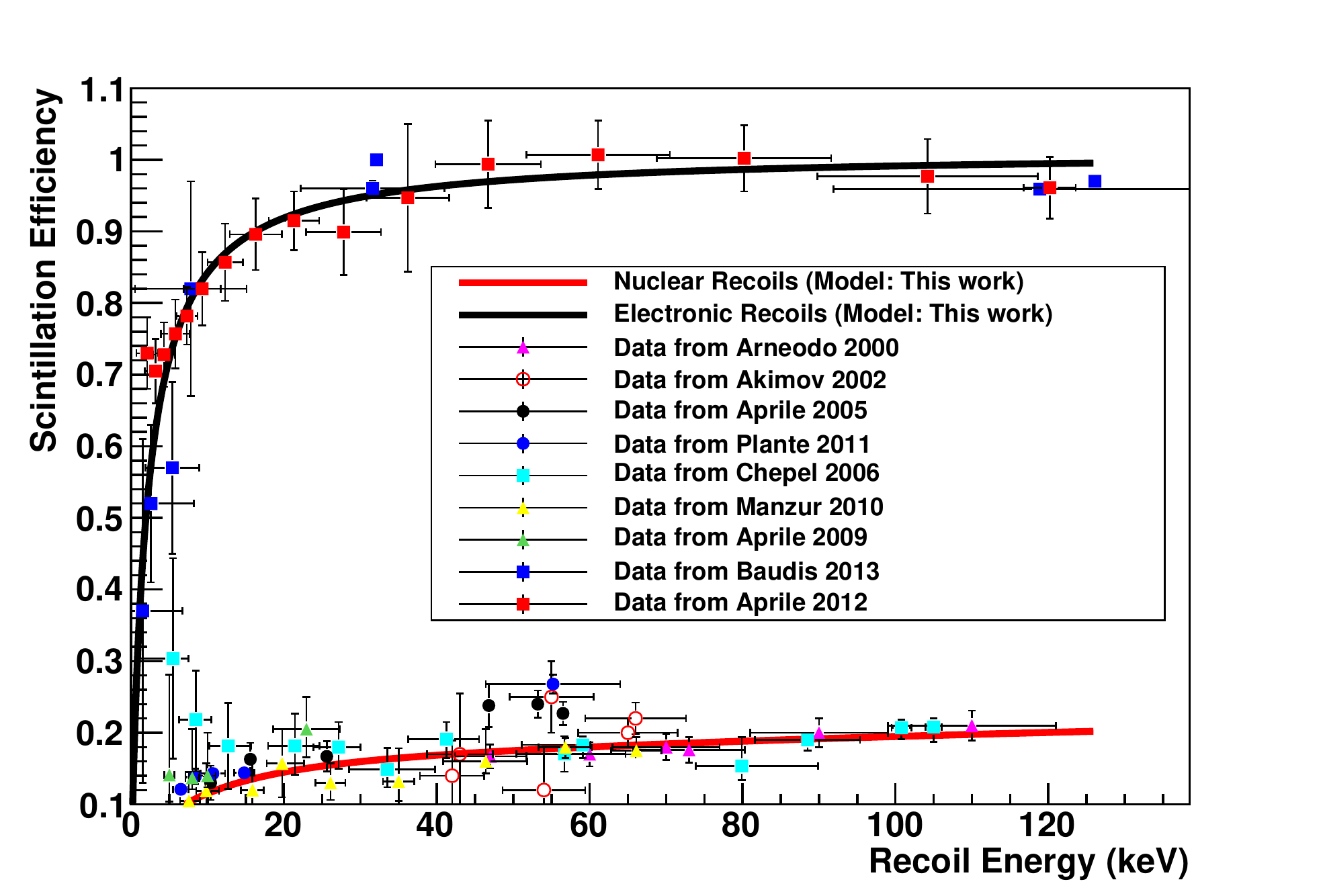}
\caption{\small{All measurements of the scintillation efficiency ($\frac{n_{\gamma}}{L_{y}^{er} \times E_{0}}$) in 
comparison with the generic model expressed by Eq.~\ref{eq:number2} at zero field.
 The data in this plot are from Arneodo 2000~\cite{arne}, Akimov 2002~\cite{akim}, Aprile 2005~\cite{apri},
Plante 2011~\cite{plan}, Chepel 2006~\cite{chep}, Manzur 2010~\cite{manz}, Aprile 2009~\cite{apri9}. The word ``Model'' in the legend represents the generic model with the Wang-Mei's recombination probability from this work.
}}
\label{fig:lightzero}
\end{center}
\end{figure}

To compare the generic model (Eq.\ref{eq:ly1}) with published data, we normalize the calculated scintillation efficiency 
to an absolute light yield of $L_{y}^{er}$ = 67 photons/keV at 122 keV from $^{57}$Co. 
Figure~\ref{fig:lightzero} shows a good agreement between the calculation of Eq.~\ref{eq:number2} and the published measurements.     
Hence, the generic model of the $W_{i}$-value, the quenching factors, and recombination proposed in this work combine to produce valid results.

\section{Summary and remarks}
\label{RD}
\subsection{Non-zero field}
We summarize the parameters used in Eq.~\ref{eq:cy1} and Eq.~\ref{eq:ly1} to calculate the charge and light yield
for NRs and ERs in Table~\ref{table:sum}.
\begin{table}[htb!!]
\caption{The summary of the parameters used in  Eq.~\ref{eq:cy1} and Eq.~\ref{eq:ly1}. }
\label{table:sum}
\begin{center} 
\begin{tabular}{|c|c|c|c|c|c|c|c|}
\hline
 & $L$& $W_{i}$ &  $t_{c}$ & t$_{pa}$ & $\alpha$ & $\beta$ & $\frac{N_{ex}}{N_{i}}$ \\
NRs & Eq.~\ref{eq:lindh} & Eq.~\ref{eq:wconst} & 15 ns & 1.5 ns & 3.617$\pm$0.521 ns &  Eq.~\ref{eq:plas2}&  Eq.~\ref{eq:ratio}\\
\hline
ERs & 1.0 & 15.6 eV& 15 ns & 1.5 ns & Eq.~\ref{eq:plas1} & Eq.~\ref{eq:plas2} & 0.1387 \\
\hline
\end{tabular}
\end{center}
\end{table}
Note that the parameters in Eq.~\ref{eq:plas2} for NRs are listed in Table~\ref{table:tab1}. Similarly, the parameters in Eq.~\ref{eq:plas1} and Eq.~\ref{eq:plas2} are summarized in Table~\ref{table:tab2}. The parameters used in Eq.~\ref{eq:lindh} and Eq.~\ref{eq:wconst} are described in subsection 2.2.1.

\begin{figure}[htb!!]
\begin{center}
\includegraphics[angle=0,width=12.cm] {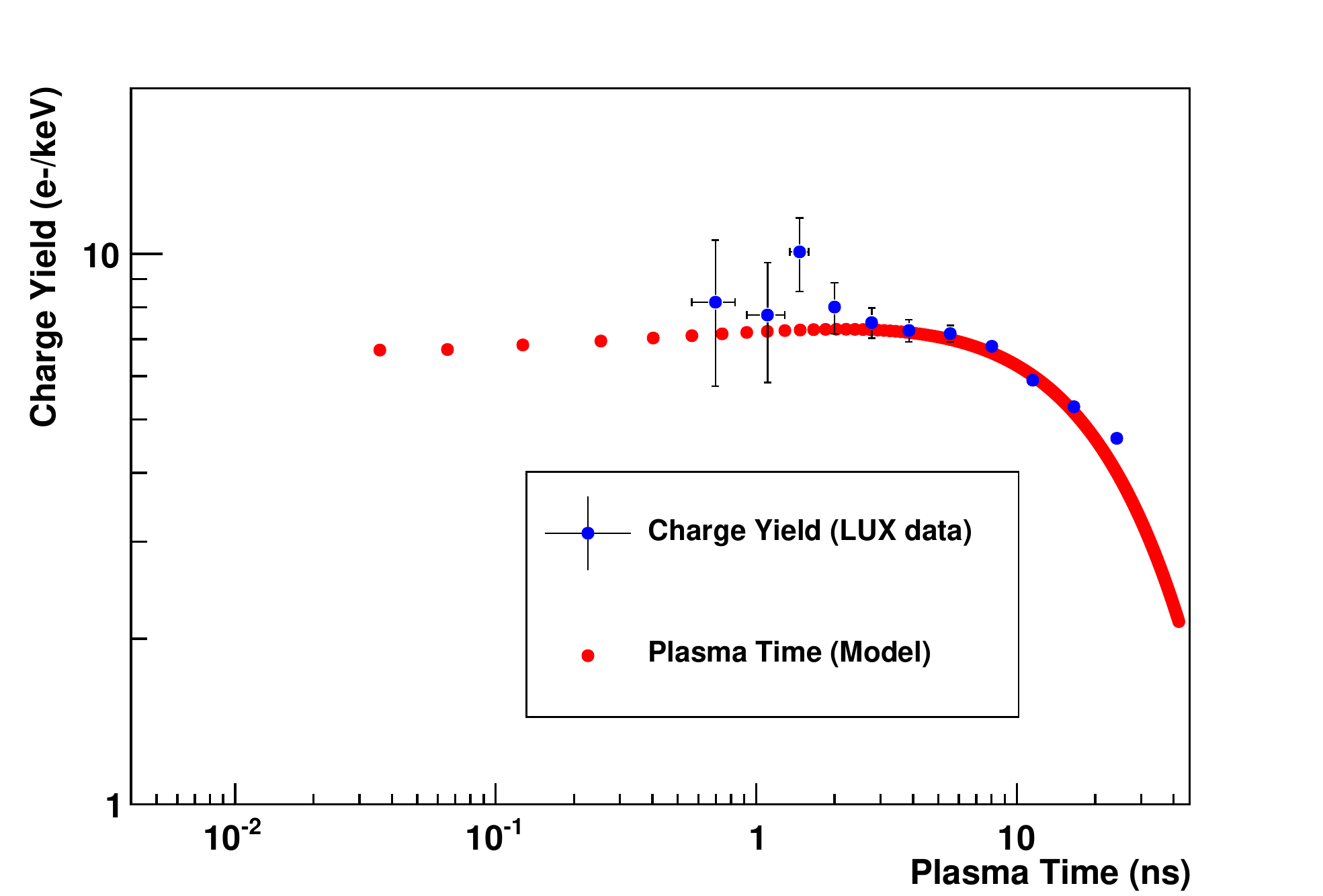}
\caption{\small{Shown is the correlation between charge yield and plasma time for NRs in comparison with LUX D-D data~\cite{Akerib}.
The blue dots are from the LUX D-D data and the red dots are calculated plasma time using Eq.~\ref{eq:plas}. }}
\label{fig:plasmavschargeyield}
\end{center}
\end{figure}

As we have shown in Section 3, in addition to the Lindhard quenching factor, 
the charge and light yield under an electric field is mainly driven by the recombination probability.
The parent recombination probability is about 6.7\% for NRs and ERs determined by the parent recombination time of 1.5 ns.   
On the one hand,  It is seen that
the plasma time plays an important role in the recombination for NRs as shown in Figure~\ref{fig:plasmavschargeyield}, which illustrates that 
the charge yield is inversely proportional to the plasma time.
On the other hand, the plasma time for ERs changes with different electric fields as shown in Figure~\ref{fig:plasField}
in Section 2.3.2. This indicates that the recombination probability for ERs slightly decreases at high electric fields.
While  at low electric fields,
the plasma time varies as a function of recoil energy. Hence the the recombination probability depends
strongly on recoil energy. 

\begin{figure}[htb!!]
\begin{center}
\includegraphics[angle=0,width=12.cm] {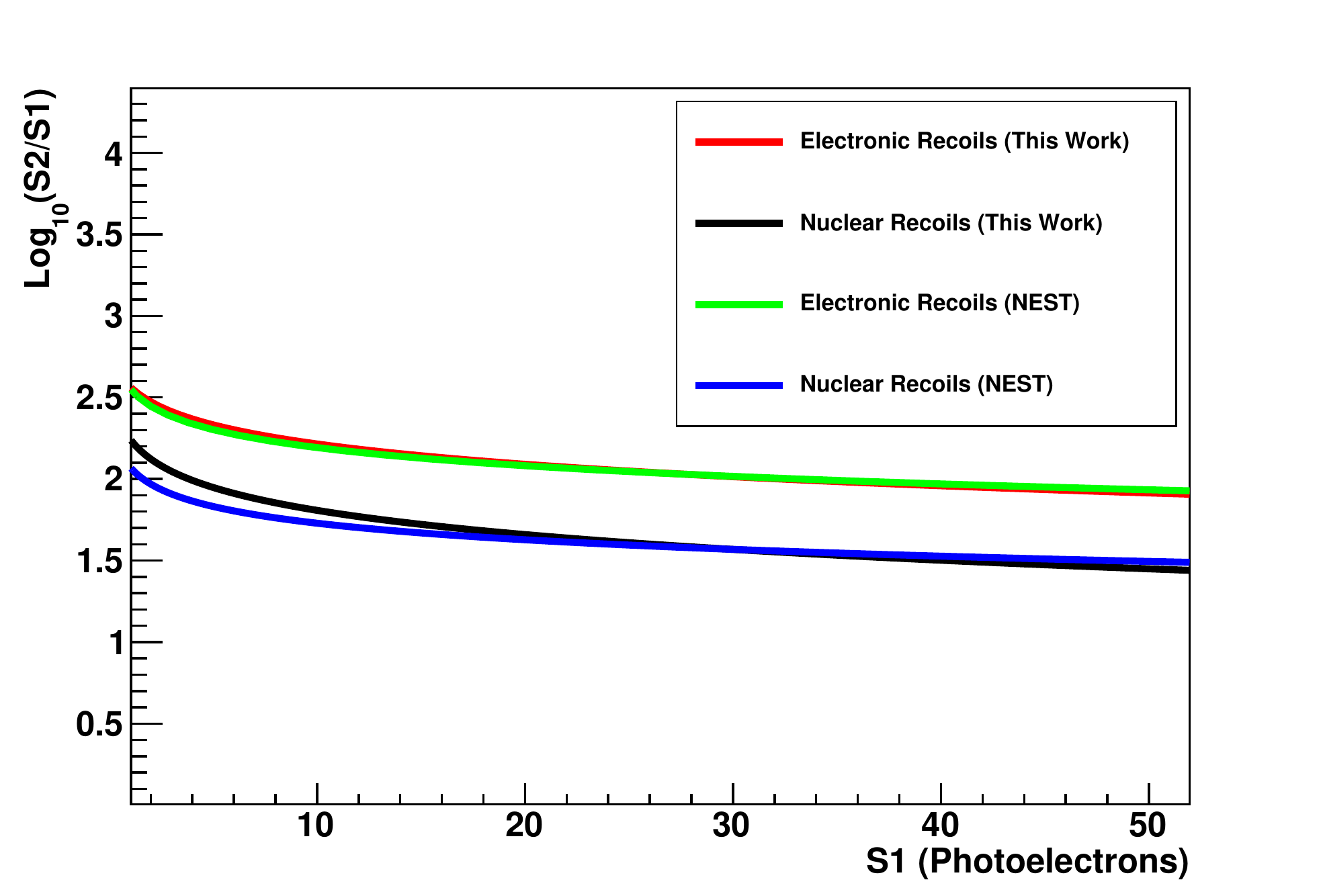}
\caption{\small{Shown is the predicted separation between NRs and ERs using the ratio of S2 to S1. The gain
factors $g_{1}$ = 0.14 and $g_{2}$ = 24.55, from LUX~\cite{LUX}, are used with a field of 180 V/cm to generate the plot. 
Note that the parameters used in the NEST model are the best-fit parameters obtained with the LUX DD neutron data and the LUX tritium data.}}
\label{fig:s2tos1}
\end{center}
\end{figure}

\begin{figure}[htb!!]
\begin{center}
\includegraphics[angle=0,width=12.cm] {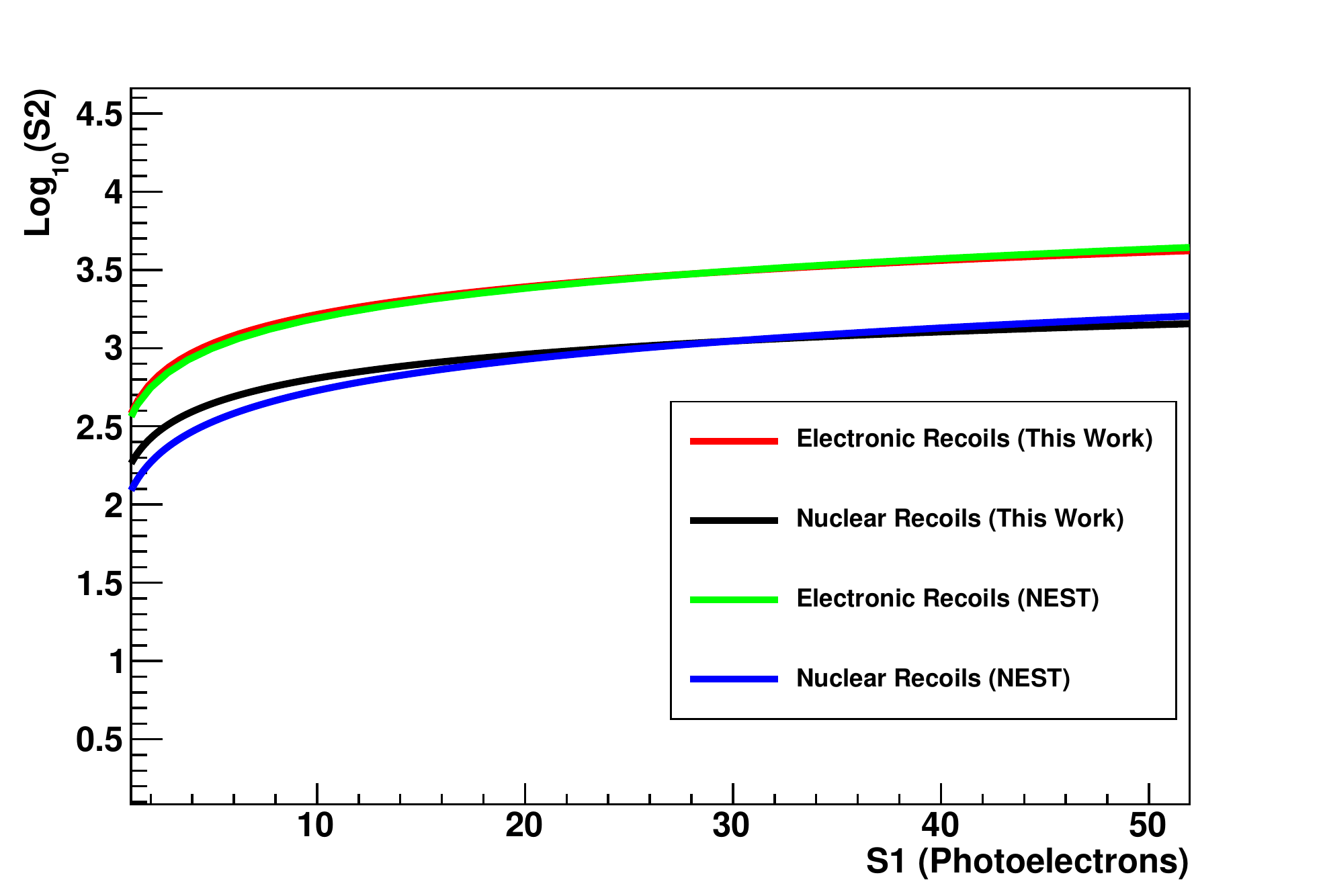}
\caption{\small{Shown is the predicted separation between NRs and ERs using $Log_{10}(S2)$ versus $S1$ analysis.
The gain
factors $g_{1}$ = 0.14 and $g_{2}$ = 24.55, from LUX~\cite{LUX}, are used with a field of 180 V/cm to generate the plot.
Note that the parameters used in the NEST model are the best-fit parameters obtained with the LUX DD neutron data and the LUX tritium data.}}
\label{fig:s2only}
\end{center}
\end{figure}

 As demonstrated by LUX, XENON100, and XENON10~\cite{LUX, xenon100, xe10}, the ratio of S2 to S1 can be used to 
discriminate NRs and ERs, as predicted by both our model and NEST as shown in Figure~\ref{fig:s2tos1}. 
One can also use $Log_{10}(S2)$ versus $S1$ to discriminate NRs from ERs as 
illustrated in Figure~\ref{fig:s2only} with an electric field of 180 V/cm.
The limitation of the separation between NRs and ERs would be only constrained by the energy resolution~\cite{mei1} in
the region of interest.
 Note that the prediction from our model and NEST, as shown in Figure~\ref{fig:s2only},
 agrees with the recent LUX analysis that has improved the sensitivity of the WIMP searches~\cite{Akerib}.


Since the difference of the plasma time between NRs and ERs is up to 10$-$50 ns at a high electric field (4000 V/cm), 
shown in Figure~\ref{fig:plastime},
this would imply a large difference in charge and light yield between ERs and NRs. 
As a result,  the charge yield from ERs at a high field (4000 V/cm) 
is a constant at a level of $\sim$ 50 $e^-/keV$ 
while the charge yield from NRs varies as a function of recoil energy, at a level of less than 10 $e^-/keV$. 
Such a large difference could enhance the separation 
between NRs and ERs. 
In theory, the minimum external field required to drift electrons away is $F$ = $\frac{v_{s}}{\mu}$, where $v_{s}$ is the speed
of sound and $\mu$ is electron mobility. For liquid xenon, $F$ $\geq$ 34.2 V/cm.
  
\subsection{Zero field}
We summarize the parameters used in Eq.~\ref{eq:ly1} to calculate the light yield
for NRs and ERs in Table~\ref{table:sum1}.
\begin{table}[htb!!]
\caption{The summary of the parameters used in Eq.~\ref{eq:ly1}. }
\label{table:sum1}
\begin{center} 
\begin{tabular}{|c|c|c|c|c|c|c|c|}
\hline
 & $L$& $W_{i}$ &  $t_{c}$ & t$_{pa}$ & $\alpha$ & $\beta$ & $\frac{N_{ex}}{N_{i}}$ \\
NRs & Eq.~\ref{eq:lindh}$\times$0.68 & Eq.~\ref{eq:wconst} & 15 ns & 1.5 ns & 14.712$\pm$1.14 ns &  2.444$\pm$0.132&  Eq.~\ref{eq:ratio}\\
\hline
ERs & Eq.~\ref{eq:photon} & 15.6 eV& 15 ns & 1.5 ns & (4.15$\pm$0.12)$\times$10$^{3}$ & 3.444$\pm$0.101 & 0.1387 \\
\hline
\end{tabular}
\end{center}
\end{table}
Note that the parameters used in Eq.~\ref{eq:lindh} and Eq.~\ref{eq:wconst} are described in subsection 2.2.1 and
the parameters used in Eq.~\ref{eq:photon} are defined in subsection 3.2.

In the case of zero field, the light yield for NRs is governed by both recombination and quenching factors. As can 
be seen in Figure~\ref{fig:compareRec}, the recombination probability decreases as recoil energy decreases in
 the energy region below 30 keV for NRs, and electrons escape recombination with xenon ions. These 
escaping electrons may be neutralised by hitting non-xenon detector components, which will not contribute to the light yield in xenon.
When the nuclear recoil energy is greater than 30 keV, the recombination probability becomes 1. The reduction of scintillation
efficiency is then solely governed by the quenching factor. 
For ERs, the recombination probability is 1 across
the entire recoil energy of interest due to the volume recombination time being greater than 10 $\mu$s. The reduction of scintillation efficiency
is solely due to the scintillation quenching.
It is worth mentioning that the e-ion pairs produced by NRs take a shorter time
to recombine (less than 200 ns) while e-ion pairs generated by ERs take a much longer time to recombine (greater than 10 $\mu$s).

\section{Conclusions}
We report a coherent and comprehensive model of light and charge yield by calculating the average energy expended per e-ion pair, 
the quenching factors, and the recombination probability 
 in liquid xenon for both non-zero field and zero field.  
As a result, 
we show that the calculations of n$_{\gamma}$ and n$_{e}$ using 
the energy-dependent $W_{i}$-value,  quenching factors,
and the field-dependent recombination probability 
agree with the available experimental results. 
A recombination model developed in this work explains the recombination process in liquid xenon, 
and predicts the dependence on recoil type, electric field, and recoil energy in the recombination probability.
At non-zero field, the charge and light yield for NRs are dominated by both recombination and Lindhard quenching factor.
While the charge and light yield for ERs are governed by recombination only. At zero field, the light yield for NRs is
affected by the quenching factors through Lindhard and Hitachi mechanisms. In the case of ERs, the e-ion recombination 
is 100\% and the light yield is
reduced by the scintillation quenching following Birks' law. 
We find that the plasma time is a main factor that governs the charge and light yield to discriminate NRs from ERs.

\section*{Acknowledgments}
The authors wish to thank Christina Keller and Amy Roberts for their careful reading of this manuscript. Additionally,
the authors would like to thank Dan McKinsey for his comments and suggestions.  
This work was supported in part by NSF PHY-0919278, NSF PHY-1242640, DOE grant DE-FG02-10ER46709, 
the Office of Research at the University of South Dakota and a research center supported by the State of South Dakota. 

%
%

\end{document}